\documentclass[reqno,centertags]{amsart}
\usepackage{amssymb}
\usepackage{upref}
\newcommand{\bbN}{{\mathbb{N}}}
\newcommand{\bbR}{{\mathbb{R}}}

\newcommand{\bbZ}{{\mathbb{Z}}}
\newcommand{\bbC}{{\mathbb{C}}}

\newcommand{\calC}{{\mathcal C}}

\newcommand{\calD}{{\mathcal D}}

\newcommand{\calM}{{\mathcal M}}
\newcommand{\calE}{{\mathcal E}}
\newcommand{\calL}{{\mathcal L}}
\newcommand{\calF}{{\mathcal F}}
\newcommand{\calK}{{\mathcal K}}

\newcommand{\calU}{{\mathcal U}}

\newcommand{\N}{n}
\newcommand{\g}{n}
\newcommand{\dott}{\,\cdot\,}
\newcommand{\hatt}{\widehat}  
\newcommand{\Pinf}{P_\infty}
\newcommand{\Div}{\operatorname{Div}}
\newcommand{\mini}{\wedge}
\newcommand{\maxi}{\vee}
\newcommand{\no}{\nonumber}
\newcommand{\lb}{\label}
\newcommand{\f}{\frac}
\newcommand{\ul}{\underline}

\newcommand{\ti}{\widetilde} 
\newcommand{\Oh}{O}

\newcommand{\bi}{\bibitem}

\renewcommand{\Im}{\text{\rm Im}}

\DeclareMathOperator{\sG}{sGmKdV}
\DeclareMathOperator{\sineG}{sG}

\DeclareMathOperator{\mKdV}{mKdV}

\allowdisplaybreaks
\numberwithin{equation}{section}

\newtheorem{theorem}{Theorem}[section]
\newtheorem{lemma}[theorem]{Lemma}

\theoremstyle{definition}

\theoremstyle{remark}
\newtheorem{remark}[theorem]{Remark}

\begin{document}
\title[sGmKdV hierarchy]{A combined sine-Gordon
and modified Korteweg--de Vries hierarchy and its
algebro-geometric solutions}
\author[Gesztesy and Holden]{Fritz Gesztesy and Helge
Holden}
\address{Department of Mathematics,
University of Missouri,
Columbia, MO 65211, USA}
\email{fritz@math.missouri.edu}
\urladdr{http://www.math.missouri.edu/people/faculty/fgesztesypt.html}
\address{Department of Mathematical Sciences,
Norwegian University of
Science and Technology, N--7034 Trondheim, Norway}
\email{holden@math.ntnu.no}
\urladdr{http://www.math.ntnu.no/\~{}holden/}

\thanks{Research supported in part by the
US National Science Foundation under Grant No.
DMS-9623121 and the Norwegian Research Council.}
\date{July 2, 1997}

\begin{abstract}
We derive a zero-curvature formalism for a combined sine-Gordon
(sG) and modified Korteweg--de Vries (mKdV) equation which
yields a local
sGmKdV hierarchy. In complete analogy to other completely
integrable hierarchies of soliton equations, such as the KdV,
AKNS, and
Toda hierarchies, the sGmKdV hierarchy is recursively
constructed by means of a fundamental polynomial formalism
involving a spectral parameter. We further illustrate our
approach by developing the basic algebro-geometric setting
for the sGmKdV hierarchy, including Baker--Akhiezer functions,
trace formulas, Dubrovin-type equations, and theta
function representations for its algebro-geometric
solutions. Although we mainly focus on sG-type equations,
our formalism also yields the
sinh-Gordon, elliptic sine-Gordon, elliptic sinh-Gordon,
and Liouville-type equations combined with the mKdV hierarchy.
\end{abstract}

\maketitle
\section{Introduction}\lb{s1}

This paper is concerned with two main issues, a systematic
derivation of a local hierarchy of nonlinear
evolution equations by embedding the sine-Gordon (sG) equation
into the modified Korteweg--de Vries
(mKdV)  hierarchy, resulting in what we call the sGmKdV hierarchy,
and a simplified approach to its algebro-geometric solutions.

A careful investigation of the (enormous) literature on
the sG equation (in light-cone coordinates)
\begin{equation}
u_{xt} = \sin(u), \lb{1.1}
\end{equation}
reveals the fact that relatively little effort
has been spent on deriving solutions which simultaneously
satisfy the whole hierarchy of sine-Gordon equations.
More significantly, the generally accepted
hierarchy in the sine-Gordon case, as originally derived by
Sasaki and Bullough \cite{SB80}, \cite{SB81} in 1980,
in sharp
contrast to other hierarchies of soliton equations such as
the KdV, AKNS, Toda, and Gelfand--Dickey hierarchies,
appears to be nonlocal in $u$ for all but the first
element \eqref{1.1} in the hierarchy,
although attempts at deriving a local sine-Gordon hierarchy
(which, however, fell short of providing an explicit
formalism) were made by S. J. Al'ber and M. S. Al'ber
\cite{AA87} in 1987.
In particular,
algebro-geometric (or periodic) solutions and their
theta function representations are not derived
for higher-order sG equations. The
current paper focuses on the close connection between the
sG equation and the mKdV hierarchy. We offer an
elementary recursive approach to a local
hierarchy which combines the sG equation and the mKdV
hierarchy in a completely integrable manner
(and similarly for the sinh-Gordon, elliptic sine-Gordon,
elliptic sinh-Gordon, and Liouville-type equations,
etc.), in the
spirit of previous treatments of the Toda
\cite{BGHT98}, Boussinesq \cite{DGU98},
AKNS \cite{GR96},
and  KdV \cite{GRT96} hierarchies, respectively. Since, in a
sense to be made precise at the end of Section \ref{s2}, the
new hierarchy embeds the sG equation into the mKdV hierarchy,
we call it the sGmKdV herarchy.

Our second major aim is to provide a simplified derivation
of the algebro-geometric solutions of the sG equation and
simultaneously
of the sGmKdV hierarchy. Due to the  intensive literature
devoted to algebro-geometric solutions of the sG equation
(and its close relatives), see, for instance
\cite{BBEIM94}, \cite{Ch78}--\cite{Da82},
\cite{Du82a}--\cite{DKN90}, \cite{DN82a}, \cite{DN82b},
\cite{Er89}--\cite{EFMM87}, \cite{FM82}, \cite{FM83},
\cite{HW93}, \cite{Ko89}, \cite{KK76}, \cite{LT88},
\cite{Mc81}, \cite{No85}, \cite{No96}, \cite{Ta90a},
\cite{TCL87},
\cite{TTCL84}, this second part of our paper might
need some justification. Our reasons for including a
detailed treatment of this topic are twofold. First of all,
it is natural to test whether the sGmKdV hierarchy
permits algebro-geometric solutions in a manner
similar to all other known hierarchies of integrable
equations mentioned previously. Secondly, we are going to
present a fundamental polynomial formalism which
considerably simplifies the existing approach to
algebro-geometric solutions even in the special case of the
original sG equation \eqref{1.1}.

Before sketching the content of each section, it seems
appropriate to point out the enormous popularity of the sG
equations and its close allies (the sinh-Gordon,
elliptic sine-Gordon, elliptic sinh-Gordon equations)
in a great variety of diverse fields. Explicit examples are
the study of surfaces with
constant negative curvature, or integrable surfaces (see,
e.g.,  \cite{Bo91a}, \cite{Bo91b},
\cite{Ei09},
\cite{EKT93}, \cite{Ko94}, \cite{MS93}, \cite{PS89}),
which points to
its actual historical origin in the 19th century,
elementary particle physics (cf.\ \cite{DEGM82},
Sects.\ 7.1--7.5), quantum optics (see \cite{DEGM82},
Sect.\ 7.8), Josephson junctions (cf.\ \cite{DEGM82},
Sect.\ 7.8.1),
nonlinear excitations in condensed matter physics
(see \cite{BK88}, \cite{BK89}), vortex structures in fluids and
plasmas (cf.\ \cite{TCL87}), and in connection with the
reduction process of Abelian integrals on hyperelliptic
curves to elliptic functions (see \cite{Ba85},
\cite{BBM86}, \cite{BBEIM94}, Sect.\ 7.9,
\cite{BBME86}--\cite{Bo84}, \cite{Sm91}--\cite{Ta90b}).
In particular, it represents one of the
celebrated examples of an infinite-dimensional completely
integrable Hamiltonian system with associated action-angle
variables in the scattering case (cf.\ \cite{FT87}, Part II,
Sects.\ II.6 and II.7, \cite{NMPZ84}, Sect.\ I.11). Moreover,
the dimensional reductions of the self-dual Yang-Mills
equations to the (elliptic) sG equation (see, e.g.,
\cite{Uh92}, \cite{Wa85}) should also be mentioned in this
connection.

In Section \ref{s2} we describe our zero-curvature
formalism for the sGmKdV hierarchy. Following a recursive
polynomial approach originally developed by Al'ber in the
context of the KdV hierarchy \cite{Al79}, \cite{Al81} and
further developed in \cite{BGHT98}, \cite{GR96}, and
\cite{GRT96} in connection with the Toda, AKNS, and KdV
hierarchies, respectively, we derive a hierarchy of local
nonlinear evolution equations whose first element is the
sG equation \eqref{1.1}. Section
\ref{s3} is devoted to a detailed study of the stationary
sGmKdV hierarchy. Following a device originally due to Jacobi
\cite{Ja46} and applied to the KdV equation by Mumford
\cite{Mu84} and McKean \cite{Mc85}, we employ the recursive
polynomial formalism of Section \ref{s2} to describe
positive divisors of degree $\N$ on a hyperelliptic curve
$\calK_\N$ of genus $\N$ associated with the $\N$th
equation in the stationary sGmKdV hierarchy. By means of a
fundamental meromorphic function $\phi$ on $\calK_\N$ we
then proceed to derive the theta function representations
of the associated Baker--Akhiezer function and all
stationary solutions of the sGmKdV hierarchy. In addition, we
consider Dubrovin-type equations for auxiliary divisors of
degree $\N$ and the corresponding trace formulas for $u(x)$
in terms of these divisors. While mKdV curves (more
generally, AKNS curves) are Toda-like curves and hence not
branched at infinity, the sG curve is KdV-like with a branch
point at infinity. In our combined sGmKdV hierarchy (for
$\alpha,\beta\neq 0$) the sG part dominates, resulting
again in a KdV-like curve branched at infinity. The
corresponding time-dependent formalism is presented in detail in Section
\ref{s4}. Appendix \ref{A} collects relevant material on
hyperelliptic curves and their theta functions.
Appendix \ref{B} sketches a new connection between our
polynomial
recursion formalism and symmetric functions related to
auxiliary divisors due to \cite{EGHL97}, which provides
new insight into the linearization of the sGmKdV
flows in Theorems \ref{theorem3.3} and \ref{theorem4.11}
and into the construction of integrable hierarchies of
soliton equations. Finally, Appendix
\ref{C} proves solvability of the principal recursion
relations
in Section \ref{s2}.

\section{The sGmKdV hierarchy} \lb{s2}

In this section we provide the basic construction of a
local sGmKdV hierarchy, embedding the sG equation into the
mKdV hierarchy in an integrable manner, using a
zero-curvature formalism.
We follow some ideas of Al'ber \cite{Al79}, \cite{Al81}, who
developed a
recursive polynomial approach for the KdV hierarchy, and
recent extensions of this formalism to the Toda,
Boussinesq, AKNS,
and KdV hierarchies in \cite{BGHT98}, \cite{DGU98},
\cite{GR96},
and \cite{GRT96}, respectively.

Assuming $u\in C^{\infty}(\bbR)$ (or meromorphic on
$\bbC$)
and $z\in\bbC$, we introduce the $2\times2$ matrix
$U(z,x)$ by
\begin{equation}
U(z,x)=-i\begin{pmatrix}\f12 u_x(x) & 1 \\ z & -\f12
u_x(x)\end{pmatrix}, \quad x\in\bbR,
\lb{2.1}
\end{equation}
and for each $\N\in\bbN_0$ the following $2\times2$  matrix
$V_\N(z,x)$
\begin{equation}
V_\N(z,x) =\begin{pmatrix} -G_{\N-1}(z,x) &
\f{1}{z}F_\N(z,x) \\
H_\N(z,x) & G_{\N-1}(z,x)\end{pmatrix}, \quad x\in\bbR,
\lb{2.2}
\end{equation}
supposing $F_\N, G_{\N-1}$, and $H_\N$ to be entire with
respect to the spectral parameter $z$ and $C^{\infty}$ (or
meromorphic) in $x$.

Postulating the stationary zero-curvature condition
\begin{equation}
V_{\N,x}=[U,V_\N] \lb{2.3}
\end{equation}
($[\dott,\dott]$ the commutator), equation \eqref{2.3} yields
the following fundamental
relationships  between the functions $F_\N$, $H_\N$ and
$G_{\N-1}$,
\begin{subequations} \lb{2.4}
\begin{align}
F_{\N,x}(z,x) & = -i u_x(x) F_\N(z,x)-2iz G_{\N-1}(z,x),
\lb{2.4a}\\
H_{\N,x}(z,x) & = i u_x(x) H_\N(z,x)+2iz G_{\N-1}(z,x),
\lb{2.4b}\\
G_{\N-1,x}(z,x) & = i(H_\N(z,x)-F_\N(z,x)). \lb{2.4c}
\end{align}
\end{subequations}
 From \eqref{2.4} one infers that
\begin{equation}
\f{d}{dx}\bigg(zG_{\N-1}(z,x)^2+F_\N(z,x) H_\N(z,x)\bigg)=0,
\lb{2.5}
\end{equation}
and hence
\begin{equation}
zG_{\N-1}(z,x)^2+F_\N(z,x) H_\N(z,x)=P_{2\N}(z), \lb{2.6}
\end{equation}
where $P_{2\N}(z)$ is $x$-independent.  It turns out that
it is more
convenient to define  $R_{2\N+1}(z)=zP_{2\N}(z)$ so that
\eqref{2.6} becomes
\begin{equation}
z^2G_{\N-1}(z,x)^2+zF_\N(z,x) H_\N(z,x)=R_{2\N+1}(z).
\lb{2.7}
\end{equation}
Using equations \eqref{2.4} one can derive individual
differential equations for $F_\N$
and
$H_\N$ as follows.  From \eqref{2.4a} and \eqref{2.4c} one infers
\begin{equation}
F_{\N,xx}=-iu_{xx}F_\N+2z(H_\N-F_\N)-u_x^2 F_\N-2z u_x G_{\N-1}.
\lb{2.8}
\end{equation}
Multiplying \eqref{2.8} by $F_\N$  one can eliminate
$G_{\N-1}$ to find
\begin{equation}
F_\N F_{\N,xx}-\f12 F_{\N,x}^2+(2z+\f12 u_x^2+i u_{xx})F_\N^2=2z
P_{2\N}=2R_{2\N+1}, \lb{2.9}
\end{equation}
and differentiating with respect to $x$ finally yields
\begin{equation}
F_{\N,xxx}+(4z+u_x^2+2i u_{xx}) F_{\N,x}+(u_x u_{xx}+
i u_{xxx})F_\N=0.
\lb{2.10}
\end{equation}
A similar analysis for $H_\N$ results in
\begin{equation}
H_\N H_{\N,xx}-\f12 H_{\N,x}^2+(2z+\f12 u_x^2-iu_{xx})H_\N^2
=2R_{2\N+1}
\lb{2.10a}
\end{equation}
and
\begin{equation}
H_{\N,xxx}+(4z+u_x^2-2i u_{xx}) H_{\N,x}+(u_x u_{xx}-
i u_{xxx})H_\N=0.
\lb{2.11}
\end{equation}
Introducing
\begin{equation}
w_\pm=-\f14(u_x^2\pm2i u_{xx}), \lb{2.12}
\end{equation}
equations \eqref{2.10} and \eqref{2.11} take the form
\begin{subequations} \lb{2.13}
\begin{align}
-\frac{1}{4}F_{\N,xxx}+(w_+ -z) F_{\N,x}+
\frac{1}{2}w_{+,x}F_\N&=0,
\lb{2.13a} \\
-\frac{1}{4}H_{\N,xxx}+(w_- -z) H_{\N,x}+
\frac{1}{2}w_{-,x}H_\N&=0,
\lb{2.13b}
\end{align}
\end{subequations}
which are identical to the corresponding equations
for the KdV
hierarchy (see
\cite{GRT96}) with potential $V=w_\pm$. Analogous assertions
apply to \eqref{2.9} and \eqref{2.10a}. Intimate connections
between
the KdV, mKdV, and sine-Gordon (resp.\ sinh-Gordon) equations
involving
 Miura-type transformations
have been known for a long time, see, for instance,
\cite{DS85},
\cite{Gu86} and more recently, \cite{TW96}. Additional
comments in this direction will be made in
Remark~\ref{remark2.3}.
Thus by making the following polynomial ansatz
with respect to the spectral parameter $z$,
\begin{equation}
F_\N(z,x)=\sum_{j=0}^\N f_{\N-j}(x) z^j, \quad
H_\N(z,x)=\sum_{j=0}^\N h_{\N-j}(x) z^j,
\lb{2.14}
\end{equation}
we first deduce by taking $z=0$ in  \eqref{2.4a} and
\eqref{2.4b} that
\begin{subequations} \lb{2.15A}
\begin{align}
f_{\N,x}(x) & =-iu_x(x) f_\N(x), \quad \N\in\bbN_0, \lb{2.15b} \\
h_{\N,x}(x) & =iu_x(x) h_\N(x), \quad \N\in\bbN_0. \lb{2.15d}
\end{align}
\end{subequations}
For $\N\in\bbN$ we conclude by combining \eqref{2.13} and
\eqref{2.14} (cf.\ \cite{GRT96})
\begin{subequations} \lb{2.15}
\begin{align}
f_0(x) & = 1, \quad \N\in\bbN, \no \\
f_{j,x}(x)&  = - \frac14 f_{j-1, xxx}(x) + w_+(x) f_{j-1,x}(x)
+ \frac12 w_{+,x}(x) f_{j-1}(x),  \lb{2.15a} \\
& \hspace*{5cm} j=1,\dots,\N, \, \N \in\bbN, \no \\
\intertext{and}
h_0(x) &  = 1, \quad \N\in\bbN, \no \\
h_{j,x}(x) & = - \frac14 h_{j-1, xxx}(x) + w_-(x) h_{j-1,x}(x)
+ \frac12 w_{-,x}(x) h_{j-1}(x),  \lb{2.15c} \\
& \hspace*{5cm} j=1,\dots,\N, \, \N \in\bbN. \no
\end{align}
\end{subequations}
Furthermore, by taking $z=0$ in  \eqref{2.6}  we find
the constraint
\begin{equation}
f_\N(x)h_\N(x)=P_{2\N}(0). \lb{2.16c}
\end{equation}
Compatibility of  $f_\N$ and $h_\N$, $\N\in\bbN$ as defined
by \eqref{2.15A}
and \eqref{2.15} is proved in Lemma \ref{lemmaC2}.

Explicitly (for $\N$ sufficiently large),
\begin{subequations} \lb{2.16}
\begin{align}
f_0  & =  1, \no \\
f_1 & =  \tfrac{1}2 w_+ + c_1=-\tfrac18(u_x^2+2i u_{xx})
+c_1,  \no \\
f_2 & =  - \tfrac18 w_{+,xx} +\tfrac38 w_{+}^2 +
 c_1\tfrac12 w_{+} + c_2 \lb{2.16a} \\
& =  -\tfrac1{32}u_{xx}^2+\tfrac1{16}u_x u_{xxx}+\tfrac{i}{16}
u_{xxxx}+\tfrac3{128}
u_x^4+\tfrac{3i}{32}u_x^2 u_{xx}-\tfrac{c_1}8
(u_x^2+2i u_{xx})+c_2, \no\\
f_3 & = \tfrac1{32}w_{+,xxxx} -
\tfrac5{16} w_{+}w_{+,xx} - \tfrac5{32} w_{+,x}^2 +
\tfrac5{16} w_{+}^3 + c_1 (- \tfrac18 w_{+,xx}
+ \tfrac38 w_{+}^2 )\no\\
& \quad +c_2 \tfrac12 w_{+} + c_3, \text{  etc.,} \no \\
\intertext{and}
h_0 & = 1, \no \\
h_1 & = \tfrac{1}2 w_- + c_1=-\tfrac18(u_x^2-2i u_{xx})
+c_1,  \no \\
h_2 & = - \tfrac18 w_{-,xx} +\tfrac38 w_{-}^2 +
 c_1\tfrac12 w_{-} + c_2 \lb{2.16b} \\
& = -\tfrac1{32}u_{xx}^2+\tfrac1{16}u_x
u_{xxx}-\tfrac{i}{16}u_{xxxx}+\tfrac3{128}
u_x^4-\tfrac{3i}{32}u_x^2 u_{xx}-\tfrac{c_1}8
(u_x^2-2i u_{xx})+c_2, \no\\
h_3 & = \tfrac1{32} w_{-,xxxx} -
\tfrac5{16} w_{-}w_{+,xx} - \tfrac5{32} w_{-,x}^2 +
\tfrac5{16} w_{-}^3 + c_1 (- \tfrac18 w_{-,xx}
+ \tfrac38 w_{-}^2 )\no \\
& \quad +c_2 \tfrac12 w_{-}  + c_3, \text{  etc.}\no
\end{align}
\end{subequations}
Here $\{c_\ell\}_{\ell\in\bbN}$ denote integration constants,
which for compatibility
reasons (see Lemma \ref{lemmaC1}) have to be
chosen identical in \eqref{2.16a} and \eqref{2.16b}.

Next, making the ansatz
\begin{equation}
G_{-1}(z,x)=0, \quad G_{\N-1}(z,x)
=\sum_{j=0}^{\N-1} g_{\N-1-j}(x) z^j,
\quad \N\in\bbN, \lb{2.22}
\end{equation}
equation \eqref{2.4a} yields
\begin{equation}
g_j(x)=\f{i}{2}(f_{j,x}(x)+i u_x(x) f_{j}(x)),
\quad j=0,\dots,\N-1, \, \N\in\bbN, \lb{2.23}
\end{equation}
and $f_{\N,x}+iu_x f_\N=0$, implying
\begin{equation}
f_\N(x)=\alpha e^{-iu(x)}, \quad \alpha\in\bbC,
\quad \N\in\bbN_0. \lb{2.24}
\end{equation}
Recording the first few  coefficients in \eqref{2.22}
then yields
(for $\N$ sufficiently large)
\begin{align}
g_0=&-\tfrac12 u_x, \no \\
g_1=&\tfrac1{16} u_x^3+\tfrac18 u_{xxx}-\tfrac{c_1}2 u_x,
\lb{2.25} \\
g_2=&-\tfrac3{256}u_x^5-\tfrac1{32} u_{xxxxx}-\tfrac{i}{32}
u_{xx} u_{xxx}
-\tfrac{5}{64} u_x^2 u_{xxx}-\tfrac{5}{64}
u_x u_{xx}^2 \no \\
&\quad +c_1(\tfrac1{16} u_x^3+\tfrac18 u_{xxx})-
\tfrac{c_2}2 u_x,
\text{  etc.}  \no
\end{align}
Equation \eqref{2.4b} yields
\begin{equation}
g_j(x)=\f{i}{2}(-h_{j,x}(x)+i u_x(x) h_{j}(x)),
\quad j=0,\dots,\N-1, \, \N\in\bbN, \lb{2.26}
\end{equation}
and  $h_{\N,x}-iu_x h_\N=0$, implying
\begin{equation}
h_\N(x)=\beta e^{iu(x)}, \quad \beta\in\bbC,
\quad \N\in\bbN_0. \lb{2.27}
\end{equation}
Finally, equation \eqref{2.4c} implies
\begin{equation}
g_{j,x}(x)=i (h_{j+1}(x)-f_{j+1}(x)),
\quad j=0,\dots, \N-1, \, \N\in\bbN. \lb{2.28}
\end{equation}
Compatibility of \eqref{2.23}, \eqref{2.24},
and \eqref{2.26}--\eqref{2.28} is proven in
Lemmas \ref{C1} and \ref{lemmaC2}.

For notational convenience we define
\begin{equation}
g_{-1}(x)=0. \lb{2.28k}
\end{equation}
Then
\begin{equation}
g_{\N-1,x}(x)=i(\beta e^{iu(x)}-\alpha e^{-iu(x)}),
\quad \N\in\bbN_0,\lb{2.30}
\end{equation}
defines the $\N$th stationary sGmKdV
equation, subject to the constraint (cf.\ \eqref{2.16c})
\begin{equation}
\alpha \beta = P_{2\N}(0). \lb{2.26a}
\end{equation}

We record the first few equations explicitly,

\begin{align}
(\beta e^{iu}-\alpha e^{-iu})&=0, \no \\
-\tfrac{i}{2}u_{xx}+(\beta e^{iu}-\alpha e^{-iu})&=0,
\lb{2.31} \\
\tfrac{i}{16}(u_x^3+2u_{xxx})_x
-c_1 \tfrac{i}{2} u_{xx} +(\beta e^{iu}-\alpha e^{-iu})&=0,
\text{  etc.} \no
\end{align}

In particular, for $\alpha=\beta\neq 0$, the first equation in
\eqref{2.31} yields
the stationary sine-Gordon
equation (in light-cone coordinates), that is,
\begin{equation}
\sin(u)=0. \lb{2.32}
\end{equation}
In the special case $\alpha=\beta=0$ one obtains $f_\N=h_\N=0$,
and hence the $(\N-1)$th
stationary KdV equation is satisfied for the potential $w_\pm$.
Introducing
\begin{equation}
v=\f{u_x}{2i}, \lb{2.32a}
\end{equation}
we see that $w_\pm$ and $v$ are related by the
Miura-transformation
\begin{equation}
w_\pm=v^2\pm v_x, \lb{2.32b}
\end{equation}
and we may conclude that $g_{\N-1,x}=0$ equals the $(\N-1)$th
stationary mKdV equation with solution $v$ for
$\N\in\bbN$.

For later purposes, it will be
convenient to record the homogeneous case defined by the
vanishing of all integration constants
$c_{\ell}$. Denote
\begin{align}
 \hat f_0&=f_0=1, &\hat g_0&=g_0=-\tfrac{1}{2} u_x,&
\hat h_0&=h_0=1, \quad \N\in\bbN, \no \\
 \hat{f}_j&=f_j\big|_{c_1=\cdots=c_j=0}, &
\hat{g}_j&=g_j\big|_{c_1=\cdots=c_j=0}, &
\hat{h}_j&=h_j\big|_{c_1=\cdots=c_j=0}, \lb{2.17aa} \\
& & &&  j&=1,\dots,\N-1, \, \N \geq 2, \no \\
\hat f_\N&=f_\N=\alpha e^{-iu}, & \hat h_\N&=h_\N=\beta e^{iu}, &
 \N&\in\bbN_0. \lb{2.17aab}
\end{align}

The connection between homogenous and nonhomogeneous
quantities then reads,
\begin{align}
& f_j=\sum_{\ell=0}^{j}c_{\ell}\hat{f}_{j-\ell},\quad
g_j=\sum_{\ell=0}^{j}c_{\ell}\hat{g}_{j-\ell},\quad
h_j=\sum_{\ell=0}^{j}c_{\ell}\hat{h}_{j-\ell}, \lb{2.18} \\
& \hspace*{4cm} c_0=1, \quad j=0,\dots,\N, \, \N\in\bbN_0. \no
\end{align}
Introducing
\begin{align}
\hatt F_j(z,x)&=\sum_{k=0}^j \hat f_{j-k}(x) z^k, \quad
\hatt H_j(z,x)=\sum_{k=0}^j \hat h_{j-k}(x) z^k, \quad \no \\
\hatt G_{j-1}(z,x)&=\sum_{k=0}^{j-1} \hat g_{j-1-k}(x) z^k,
\quad j=0,\dots,\N, \, \N\in\bbN_0 \lb{2.19}
\end{align}
then results in
\begin{equation}
F_\N=\sum_{j=0}^\N c_{\N-j}\hatt F_j, \quad
G_{\N-1}=\sum_{j=0}^{\N-1} c_{\N-1-j}\hatt G_j, \quad
H_\N=\sum_{j=0}^\N c_{\N-j}\hatt H_j, \quad \N\in\bbN_0,
\lb{2.20} \
\end{equation}
with $\hat f_\N=f_\N=\alpha e^{-iu}$ and
$\hat h_\N=h_\N=\beta e^{iu}$
as in \eqref{2.17aab}.

Finally, we turn to the time-dependent sGmKdV hierarchy.
Introducing a deformation parameter $t_\N\in\bbR$ into $u$ (i.e.,
$u(x) \to u(x,t_\N)$), the corresponding zero-curvature
relation
reads
\begin{equation}
U_{t_\N}-V_{\N,x}+[U,V_\N]=0, \quad \N\in\bbN_0,\lb{2.33}
\end{equation}
which results in the following set of equations
\begin{subequations}\lb{2.34}
\begin{align}
u_{xt_\N}(x,t_\N)&=-2iG_{\N-1,x}(x,t_\N)-2(H_\N(x,t_\N)-
F_\N(x,t_\N)),
\lb{2.34a} \\
F_{\N,x}(x,t_\N)&=-i u_x(x,t_\N) F_\N(x,t_\N)-
2iz G_{\N-1}(x,t_\N),
\lb{2.34b} \\
H_{\N,x}(x,t_\N)&=i u_x(x,t_\N) H_\N(x,t_\N)+
2iz G_{\N-1}(x,t_\N).  \lb{2.34c}
\end{align}
\end{subequations}
We follow a more elaborate notation inspired by Hirota's
$\tau$-function approach and indicate the individual $\N$th
sGmKdV flow by a separate time variable $t_\N\in\bbR$. (The
latter notation suggests consideration of all sGmKdV flows
simultaneously by introducing $\ul t =(t_0,t_1,t_2,\dots)$.)

Inserting the polynomial expressions for $F_\N$, $H_\N$,
and $G_{\N-1}$ into \eqref{2.34b} and  \eqref{2.34c},
respectively, first yields
\begin{equation}
f_{\N,x}(x,t_\N)  =-iu_x(x,t_\N) f_\N(x,t_\N), \quad
h_{\N,x}(x,t_\N)  =iu_x(x,t_\N) h_\N(x,t_\N),
\quad \N\in\bbN_0.
\lb{2.34kk}
\end{equation}
In the general case we find
\begin{subequations} \lb{2.34k}
\begin{align}
f_0(x,t_\N) & = 1, \quad \N\in\bbN, \lb{2.34aa} \\
f_{j,x}(x,t_\N)&  = - \frac14 f_{j-1, xxx}(x,t_\N) +
w_+(x,t_\N) f_{j-1,x}(x,t_\N)
+ \frac12 w_{+,x}(x,t_\N) f_{j-1}(x,t_\N),  \no \\
& \hspace*{6cm} j=1,\dots,\N-1, \, \N \geq 2, \no \\
\intertext{and}
h_0(x,t_\N) &  = 1, \quad \N\in\bbN, \lb{2.34ab} \\
h_{j,x}(x,t_\N) & = - \frac14 h_{j-1, xxx}(x,t_\N) +
w_-(x,t_\N) h_{j-1,x}(x,t_\N)
+ \frac12 w_{-,x}(x,t_\N) h_{j-1}(x,t_\N),  \no \\
& \hspace*{6cm} j=1,\dots,\N-1, \, \N\geq 2, \no
\end{align}
\end{subequations}
and (recall our convention \eqref{2.28k})
\begin{equation}
u_{xt_\N}(x,t_\N) =-2ig_{\N-1,x}(x,t_\N)-2(h_\N(x,t_\N)
-f_\N(x,t_\N)), \quad \N\in\bbN_0, \lb{2.35}
\end{equation}
in addition to equations \eqref{2.23}, \eqref{2.24},
\eqref{2.26},
\eqref{2.27}, and
\eqref{2.28} (the latter equation for $j=0,\dots, \N-2$
and only for $\N \geq 2$).
Varying $\N\in \bbN_0$ then defines the
time-dependent sGmKdV hierarchy by
\begin{align}
\sG_\N(u(x,t_\N))&=u_{xt_\N}(x,t_\N)+2ig_{\N-1,x}(x,t_\N)
+2(\beta e^{iu(x,t_\N)}-\alpha e^{-iu(x,t_\N)}) \no \\
&=0, \hspace*{6cm} \N\in\bbN_0. \lb{2.36}
\end{align}
Explicitly, the first few equations read
\begin{align}
\sG_0(u) & =  u_{xt_0}+2(\beta e^{iu}-\alpha e^{-iu})=0,
\no \\
\sG_1(u) & =  u_{xt_1}-iu_{xx}+2(\beta e^{iu}-
\alpha e^{-iu})=0, \lb{2.37} \\
\sG_2(u) & =  u_{xt_2}+\tfrac{i}8(u_x^3+2u_{xxx})_x
-c_1 i u_{xx} +2(\beta e^{iu}-\alpha e^{-iu})=0,
\text{  etc.} \no
\end{align}

In contrast to the stationary case, the constraint
\eqref{2.26a}
does not apply in the $t_\N$-dependent context \eqref{2.36}
(since the left-hand side of \eqref{2.34a} is nonvanishing).
Moreover, the compatibility requirements for $f_\N$ and
$h_\N$ in \eqref{2.15A} and \eqref{2.15}, which were of
great significance in
the stationary case, are absent in the
present time-dependent situation.

\begin{remark} \lb{remark2.1}
Here $\sG_0(u)=\sineG(u)=0$ is {\it the} sine-Gordon  equation in
light-cone coordinates choosing $\alpha=\beta=i/4$,
\begin{equation}
u_{xt_0}=\sin(u).\lb{2.38}
\end{equation}
We note that $u_{xt_{0}}=2(\alpha e^{-iu} -\beta e^{iu})$
reduces to $v_{xt_{0}}=\sin (v)$, where
$v(x,t_{0})=u(x/a,t_{0})  +(2i)^{-1} \ln (\beta/\alpha)$,
$a=-2i(\alpha\beta)^{1/2}$,
$\alpha,\beta\in\bbC\setminus\{0\}$.
Writing $v=iu$ and setting $\alpha=\beta=1/4$, one finds
\begin{equation}
v_{xt_0}=\sinh(v),\lb{2.39}
\end{equation}
that is, the sinh-Gordon equation. Similarly, introducing
$\xi = x+t_0$, $\eta =x-t_0$ produces the (hyperbolic)
sine-Gordon
and (hyperbolic) sinh-Gordon equations in laboratory coordinates
\begin{equation}
u_{\xi \xi}-u_{\eta \eta}=\sin (u),
\quad v_{\xi \xi}-v_{\eta \eta}=\sinh (v) \lb{2.39k}
\end{equation}
and $\xi =x+t_0$, $\tau =i(x-t_0)$ produces the elliptic
sine-Gordon
and elliptic sinh-Gordon equations
\begin{equation}
u_{\xi \xi}+u_{\tau \tau}=\sin (u),
\quad v_{\xi \xi}+v_{\tau \tau}=\sinh (v). \lb{2.39l}
\end{equation}
Similarly, the case
$\alpha=0$ yields
\begin{equation}
u_{xt_0}=-2\beta e^{iu} \lb{2.39a}
\end{equation}
and a similar equation in the case $\beta =0$. Hence writing
$v=iu$ and changing coordinates $(x,t_0)\to(\xi,\tau)$
yields the Liouville hierarchy for $v$ (cf., e.g.,
\cite{TCC86})
starting with
\begin{equation}
v_{\xi \xi} + v_{\tau \tau} = 2i\beta e^{-v}. \lb{2.39b}
\end{equation}
In particular, the results in
Sections \ref{s3} and \ref{s4} extend to these hierarchies
and hence
we omit further distinctions and focus on the sGmKdV
hierarchy in light-cone coordinates for the rest of this paper.

Finally, in the case $\alpha=\beta=0$, we define
\begin{equation}
v(x,t_\N)=\f{u_x(x,it_\N)}{2i}, \lb{2.39c}
\end{equation}
and the sGmKdV hierarchy reduces to
\begin{equation}
v_{t_\N}(x,t_\N)-ig_{\N-1,x}(x,t_\N)=0, \quad \N\in\bbN,
\lb{2.39d}
\end{equation}
which equals the $(\N-1)$th modified KdV equation with
solution $v$.
\end{remark}

\begin{remark} \lb{remark2.2}
In addition to the nonlocal sine-Gordon hierarchies constructed
by Sasaki and Bullough \cite{SB80}, \cite{SB81}, we are aware of
a few other attempts introducing a nonlocal sG hierarchy. For
instance, Newell \cite{Ne85}, Sect.\ 5k introduces a
nonlocal sG hierarchy
using an extension of the AKNS hierarchy, Tracy and Widom
\cite{TW96}
discuss the sG hierarchy in close connection with the mKdV
hierarchy,
and Gu \cite{Gu86} derives a generalized mKdV-sG
hierarchy. On the other hand, the existence of an infinite
sequence of conservation laws for \eqref{2.38} polynomial in $u$
and its $x$-derivatives (see, e.g., \cite{SK74}) suggests the
existence of a local hierarchy involving the sG equation.
Moreover, the fact that the usual
zero-curvature representation for \eqref{2.38} is gauge equivalent
to that of the nonlinear Schr\"odinger (nS) equation as
discussed in
\cite{FT87}, Part II, Sect.\ II.7 (see also \cite{Bo91a},
Sect.\ 5
and \cite{PS89}), and given the fact that the nS hierarchy
is well-known to be local,
also hints to the existence of a local hierarchy
involving the sG equation even though
no further details seem to have been worked out. Finally,
the strongest indication for the existence of such a local
hierarchy is provided by the so called
$\ul \mu$-representation
of the algebro-geometric sG solutions (cf.\ Remark
\ref{remark4.7}) as discussed in \cite{AA87},
\cite{EGHL97}, \cite{EF85}, \cite{FM82}, \cite{TCL87},
\cite{TTCL84}.
The only work we are aware of that clearly aimed at a recursive
construction of a local hierarchy involving the sG equation
is due to Al'ber and  Al'ber \cite{AA87}.
Their approach, however, seems different from ours
and does not
start from a zero-curvature representation. While a set of
recurrence relations is claimed in equations (4.4) and (4.5) of
\cite{AA87}, no explicit formulas are offered and the precise
form of the higher-order equations is not detailed.
\end{remark}

\begin{remark} \lb{remark2.3}
The relationship between the KdV and mKdV and the sGmKdV
hierarchy \eqref{2.36}, alluded to in \eqref{2.12}, \eqref{2.13},
the paragraph following \eqref{2.13}, \eqref{2.32a}, \eqref{2.32b},
\eqref{2.39c}, and \eqref{2.39d}, can be made precise as follows.
The equation
\begin{equation}
\Psi_{x}=U\Psi, \quad \Psi=\begin{pmatrix} \psi_1 \\
\psi_2\end{pmatrix}, \lb{2.54}
\end{equation}
is equivalent to
\begin{equation}
\psi_{1,xx}=(w_{+}-z)\psi_{1}, \quad
\psi_{2,xx}=(w_{-}-z)\psi_{2}, \lb{2.55}
\end{equation}
with
\begin{equation}
w_{\pm}=v^{2}\pm v_{x}, \quad v=u/(2i). \lb{2.56}
\end{equation}
Equation \eqref{2.56} represents the familiar Miura
transformation between solutions $w_{\pm}$ of the KdV
hierarchy and $v$ of the mKdV hierarchy. Since
$\sG_{n}(u)=0$ reduces to $\mKdV_{n-1}(v)=0$, $n\in\bbN$ for
$\alpha=\beta=0$ as discussed in Remark~\ref{remark2.1},
$\sG_{n}(u)=0$ for general $\alpha,\beta\in\bbC$ represents
a completely integrable linear combination of the
$\sineG (u)$ equation and the $\mKdV_{n-1}(v)$ equation. This
corresponds to our choice \eqref{2.2} of $zV_{n}(z,x)$
being a polynomial with respect to $z$. The usually
considered nonlocal sG hierarchy then corresponds to a
choice of $zV_{n}(z,x)$ being rational in $z$ with a
nontrivial principal part.
\end{remark}

\section{The stationary sGmKdV formalism} \lb{s3}
\setcounter{theorem}{0}

This section is devoted to a detailed study of the stationary
sGmKdV hierarchy and its algebro-geometric solutions. Our principal
tool will be a combination of the polynomial recursion formalism
introduced in Section \ref{s2} and a meromorphic function $\phi$
on a hyperelliptic curve $\calK_\N$ defined in terms of
$R_{2\N+1}(z)$. Moreover, we discuss in detail the associated
stationary Baker--Akhiezer vector $\Psi(z,x,x_0)$ and associated
auxiliary positive divisors of degree $\N$ on $\calK_\N$.

Throughout this section we assume \eqref{2.3} (resp.\ \eqref{2.4}),
\eqref{2.14}, \eqref{2.15}, \eqref{2.16}, and freely employ the
formalism
developed in \eqref{2.1}--\eqref{2.20}.

Returning to \eqref{2.7} we infer from \eqref{2.14} that
$R_{2\N+1}(z)=zP_{2\N}(z)$ is a
monomial in $z$ of degree $2\N+1$ of the form
\begin{equation}
R_{2\N+1}(z)=\prod_{m=0}^{2\N}(z-E_m), \quad E_0=0,\, E_1,\dots,
E_{2\N}\in\bbC. \lb{3.1}
\end{equation}
Computing
\begin{align}
\det(wI_2-iV_\N(z,x)) & =  w^2 - \det(V_\N(z,x)) \no \\
& =   w^2 + G_{\N-1}(z,x)^2
+\f{1}{z}F_\N(z,x)H_\N(z,x) \lb{3.1d} \\
& = w^2 +\f{1}{z^2}R_{2\N+1}(z) \no
\end{align}
(with $I_2$ the identity matrix in $\bbC^2$), we are led to
introduce the (possibly singular)
hyperelliptic  curve $\calK_\N$ of arithmetic genus $\N$ defined by
\begin{equation}
\calK_\N \colon \calF_\N(z,y)=y^2-R_{2\N+1}(z)=0. \lb{3.2}
\end{equation}
We compactify $\calK_\N$ by adding the point $\Pinf$, still denoting
its projective closure by  $\calK_\N$.  Hence $\calK_\N$
becomes a two-sheeted Riemann surface
of arithmetic genus $\N$.  Finite points $P$ on $\calK_\N$ are
denoted by $P=(z,y)$, where $y(P)$ denotes the meromorphic
function on $\calK_\N$ satisfying
$\calF_\N(z,y)=0$.  The complex structure on $\calK_\N$ is then
defined in a standard manner.
Furthermore,  we introduce the involution
\begin{equation}
*\colon \calK_\N\to\calK_\N, \quad P=(z,y)\mapsto P^*=(z,-y).
\lb{3.3}
\end{equation}
For further properties and notation concerning
hyperelliptic curves we refer to Appendix \ref{A}.

In order to avoid numerous case distinctions because of
the trivial
case $\N=0$, we shall assume $\N\in\bbN$ for the remainder
of this
section. Moreover, to simplify our presentation in the
following, we will subsequently focus on sG-type equations
and hence
assume
\begin{equation}
\alpha, \, \beta \in \bbC\setminus \{0\}. \lb{3.3a}
\end{equation}
By \eqref{2.26a} this is equivalent to
\begin{equation}
P_{2\N}(0)=\prod_{m=1}^{2\N}E_m =\alpha\beta \neq 0. \lb{3.3b}
\end{equation}
Hence, from this point on, we suppose
\begin{equation}
E_0=0, \quad E_m\in\bbC \setminus \{0\}, \quad m=0,\dots,2\N.
\lb{3.1a}
\end{equation}

In the following we need the roots of the polynomials
$F_\N$ and $H_\N$, and hence introduce
\begin{equation}
F_\N(z,x)=\prod_{j=1}^\N(z-\mu_j(x)), \quad
H_\N(z,x)=\prod_{j=1}^\N(z-\nu_j(x)). \lb{3.7}
\end{equation}
Next, define
\begin{subequations} \lb{3.12}
\begin{align}
\hat\mu_j(x)&=(\mu_j(x),-\mu_j(x)G_{\N-1}
(\mu_j(x),x))\in\calK_\N,
\quad j=1,\dots,\N, \lb{3.12a}\\
\hat\nu_j(x)&=(\nu_j(x),\nu_j(x)G_{\N-1}
(\nu_j(x),x))\in\calK_\N,
\quad j=1,\dots,\N \lb{3.12b}
\end{align}
\end{subequations}
and
\begin{equation}
P_0 = (0,0). \lb{3.12d}
\end{equation}
Define the fundamental meromorphic function $\phi$
on $\calK_\N$ by
\begin{equation}
\phi(P,x)=\f{y-zG_{\N-1}(z,x)}{F_\N(z,x)}
=\f{zH_{\N}(z,x)}{y+z G_{\N-1}(z,x)}, \quad
x\in\bbR, \, P=(z,y)\in\calK_\N, \lb{3.4}
\end{equation}
with divisor $(\phi(P,x))$ (cf.\ the
notation for divisors introduced in \eqref{A.17}
and \eqref{A.18})
given by
\begin{equation}
(\phi(P,x))=\calD_{P_0\hat{\ul\nu}(x)}-
\calD_{\Pinf\hat{\ul\mu}(x)}, \lb{3.13}
\end{equation}
where we abbreviated
\begin{equation}
\hat{\ul\mu}(x)=(\hat\mu_1(x),\dots,\hat\mu_\N(x)), \quad
\hat{\ul\nu}(x)=(\hat\nu_1(x),\dots,\hat\nu_\N(x)). \lb{3.14}
\end{equation}
The stationary Baker--Akhiezer function
\begin{equation}
\Psi(P,x,x_0)=\begin{pmatrix} \psi_1(P,x,x_0) \\
\psi_2(P,x,x_0)
\end{pmatrix},\lb{3.5}
\end{equation}
is defined on $\calK_\N \setminus \{\Pinf\}$ by
\begin{align}
\psi_1(P,x,x_0)&=\exp\left(-\f{i}2(u(x)-u(x_0))+
i\int_{x_0}^x dx'\,
\phi(P,x')\right),
\lb{3.6a}\\
\psi_2(P,x,x_0)&=-\psi_1(P,x,x_0) \phi(P,x), \lb{3.6b} \\
& \hspace*{3cm} (x,x_0)\in\bbR^2, \,
P\in\calK_\N \setminus \{\Pinf\}. \no
\end{align}
We summarize the fundamental properties of $\phi$
and $\Psi$
in the
following lemma.
\begin{lemma} \lb{lemma3.1}
Assume \eqref{3.1a}, $P=(z,y)\in\calK_\N
\setminus \{\Pinf\}$,
and let
$(z,x,x_0)\in\bbC \times \bbR^2$. Then $\phi$ satisfies
\begin{align}
&\phi(P,x)^2-i\phi_x(P,x)=z+u_x(x)\phi(P,x), \lb{3.8a}\\
&\phi(P,x)+\phi(P^*,x)=-2z\f{G_{\N-1}(z,x)}{F_\N(z,x)},
\lb{3.8b}\\
&\phi(P,x)\phi(P^*,x)=-z\f{H_{\N}(z,x)}{F_\N(z,x)},
\lb{3.8c}\\
&\phi(P,x)-\phi(P^*,x)=2\f{y(P)}{F_\N(z,x)}, \lb{3.8d}
\end{align}
while $\Psi$ is meromorphic on $\calK_\N\setminus\{\Pinf\}$
and fulfills
\begin{align}
&\Psi_x(P,x,x_0)=U(z,x)\Psi(P,x,x_0), \lb{3.9a} \\
&-\f{y(P)}{z} \Psi(P,x,x_0)=V_\N(z,x)\Psi(P,x,x_0),
\lb{3.9ab}  \\
&\psi_1(P,x,x_0)=\bigg(\f{F_\N(z,x)}{F_\N(z,x_0)}\bigg)^{1/2}
\exp\bigg(iy(P)\int_{x_0}^x \f{dx'}{F_\N(z,x')}\bigg),
\lb{3.9b} \\
&\psi_1(P,x,x_0)\psi_1(P^*,x,x_0)=\f{F_\N(z,x)}
{F_\N(z,x_0)},\lb{3.9c} \\
&\psi_2(P,x,x_0)\psi_2(P^*,x,x_0)=-z\f{H_\N(z,x)}
{F_\N(z,x_0)},\lb{3.9d} \\
&\psi_1(P,x,x_0)\psi_2(P^*,x,x_0)+
\psi_1(P^*,x,x_0)\psi_2(P,x,x_0)
=2z\f{G_{\N-1}(z,x)}{F_\N(z,x_0)}, \lb{3.9e} \\
&\psi_1(P,x,x_0)\psi_2(P^*,x,x_0)-
\psi_1(P^*,x,x_0)\psi_2(P,x,x_0)
=2\f{y(P)}{F_\N(z,x_0)}. \lb{3.9f}
\end{align}
Moreover, $u$ satisfies the following trace relations
\begin{subequations}\lb{3.10}
\begin{align}
u(x)&=i\ln\bigg((-1)^\N \alpha^{-1}
\prod_{j=1}^\N\mu_j(x) \bigg),
 \lb{3.10a}\\
u(x)&=-i\ln\bigg((-1)^\N \beta^{-1}
\prod_{j=1}^\N\nu_j(x) \bigg),
 \lb{3.10b}
\end{align}
\end{subequations}
where (cf.\ \eqref{3.3b})
\begin{equation}
\alpha\beta=\prod_{m=1}^{2\N}E_m \neq 0. \lb{3.11}
\end{equation}
\end{lemma}
\begin{proof}
Equation \eqref{3.8a} follows  using the definition
\eqref{3.4} of $\phi$
as well as
relations \eqref{2.4}. The other relations,
\eqref{3.8b}--\eqref{3.8d},
are easy consequences of $y(P^*)=-y(P)$ in addition to
\eqref{2.4}.

By \eqref{3.6a} and \eqref{3.6b}, $\Psi$ is
meromorphic away
from the poles $\hat \mu_j(x')$ of $\phi(\dott,x')$. By
\eqref{2.4a}, \eqref{3.4}, and \eqref{3.12a},
\begin{equation}
i\phi(P,x') \underset{P\to\hat \mu_j(x')}{=}
-\frac{d}{dx'} \ln
(F_\N(z,x'))\big|_{z=\mu_j(x')} + \Oh(1) \lb{3.24a}
\end{equation}
and hence $\psi_1$ is meromorphic on
$\calK_\N\setminus\{\Pinf\}$ by
\eqref{3.6a}. Since $\phi$ is meromorphic on $\calK_\N$,
$\psi_2$ is meromorphic on $\calK_\N\setminus\{\Pinf\}$.
The remaining properties of $\Psi$
can be verified by using the  definition
\eqref{3.5}--\eqref{3.6b} as well as
the relations \eqref{3.8a}--\eqref{3.8d}.  In particular,
equation
\eqref{3.9b} follows by
inserting the definition of $\phi$, \eqref{3.4}, into
\eqref{3.6a}, using
\eqref{2.4a}.

Finally, the trace relation \eqref{3.10a} follows by
considering the
constant term in
$F_\N$ in \eqref{2.14} combined with \eqref{2.24} and
\eqref{3.7}. Equation
\eqref{3.10b}
can be deduced in a similar way studying the polynomial
$H_\N$.
\end{proof}

Next, we derive differential equations, (i.e.,
Dubrovin-type equations) for $\mu_j$ and $\nu_j$. For
this purpose,
and for the theta function representations to follow
in Theorem
\ref{theorem3.3} we need to assume that $\calK_\N$ is
nonsingular.
Summing up all our conditions on $\{E_m\}_{m=0,\dots,2\N}
\subset \bbC$ we thus
assume
\begin{equation}
E_0 = 0, \quad E_m \neq E_{n} \text{ for } m \neq n,
\, m,n=0,1, \dots, 2\N \lb{3.14a}
\end{equation}
for the rest of this section.

\begin{lemma} \lb{lemma3.2}  Assume \eqref{3.14a} and
suppose that the zeros $\{\mu_j(x) \}_{j=1,\dots,\N}$
of $F_\N(\dott,x)$
remain distinct for $x\in\Omega$, where
$\Omega\subseteq\bbR$ is an open interval.
Then $\{\mu_j(x) \}_{j=1,\dots,\N}$ satisfies the
following first-order system of differential equations
\begin{equation}
\mu_{j,x}(x)=-2i\f{y(\hat\mu_j(x))}
{\prod_{\substack{\ell=1\\ \ell\neq j}}^\N(\mu_j(x)-
\mu_\ell(x))},
\quad j=1, \dots, \N, \, x\in \Omega, \lb{3.15}
\end{equation}
with initial conditions
\begin{equation}
\hat\mu_j(x_0)\in\calK_\N,\quad j=1, \dots, \N \lb{3.16}
\end{equation}
for some fixed $x_0\in\Omega$.  The initial value
problem \eqref{3.15},
\eqref{3.16} has a unique solution
$\{\hat\mu_j(x)\}_{j=1,\dots,\N}$ such that
\begin{equation}
\hat\mu_j(x)\in C^\infty(\Omega,\calK_\N),
\quad j=1, \dots, \N. \lb{3.17}
\end{equation}
For the zeros $\{\nu_j(x) \}_{j=1,\dots,\N}$ of
$H_\N(\dott,x)$ identical
statements hold with $\mu$
replaced by $\nu$. In particular,
$\{\nu_j(x) \}_{j=1,\dots,\N}$
satisfies
\begin{equation}
\nu_{j,x}(x)=-2i\f{y(\hat\nu_j(x))}
{\prod_{\substack{\ell=1\\ \ell\neq j}}^\N
(\nu_j(x)-\nu_\ell(x))},
\quad j=1, \dots, \N, \, x\in \Omega. \lb{3.18}
\end{equation}
\end{lemma}
\begin{proof}
We only prove  equation \eqref{3.15} since the proof of
\eqref{3.18}
follows in an identical manner. Inserting
$z=\mu_j$ into  equation \eqref{2.4a}, one concludes from
\eqref{3.12a}
\begin{equation}
F_{\N,x}(\mu_j)=-\mu_{j,x}
\prod_{\substack{\ell=1\\ \ell\neq j}}^\N(\mu_j-\mu_\ell)
=-2i\mu_j G_{\N-1}(\mu_j)=2iy(\hat \mu_j), \lb{3.19}
\end{equation}
proving \eqref{3.15}. The smoothness assertion \eqref{3.17}
is clear as long as $\hat \mu_j(x)$ stays away from the
branch points $(E_m,0)$. In case $\hat \mu_j(x)$ hits such a
branch point, one can use the local chart around $(E_m,0)$
(with local coordinate $\zeta=\sigma (z-E_m)^{1/2}$,
$\sigma=\pm 1$) to verify
\eqref{3.17}.
\end{proof}

\begin{remark} \lb{remark3.2a}
If $\alpha=0$ (as in Liouville-type equations \eqref{2.39b}),
the fact that
$f_\N(x)$ $=(-1)^\N \prod_{j=1}^\N \mu_j(x) =0$ forces
(at least)
one $\hat \mu_{j_0}(x)$ to coincide with $P_0=(0,0)$
and hence to be $x$-independent. Similarly, if
$\alpha=\beta=0$, then $\hat \mu_{j_1}(x) =
\hat \nu_{j_2}(x) = P_0$ for some $j_1,j_2\in\{1,
\dots,\N\}$.
\end{remark}

We end this section by deriving representations of the
functions $\phi$, $\Psi$ as
well as $u$, in terms of the Riemann theta function
$\theta$ of $\calK_\N$. We will be relatively brief in our
exposition and refer to \cite{BBEIM94}, Chs.\ 2--4,
\cite{Du77}--\cite{Du82a}, \cite{Du83}--\cite{DMN76},
\cite{GH98}, \cite{Kr77a}, \cite{Kr77b}, \cite{Ma76},
and \cite{NMPZ84}, Ch.\ II for the general algebro-geometric
approach to integrable soliton equations. We freely use the
notation established in Appendix \ref{A}.

We choose an arbitrary but fixed base point
$Q_0$ on $\calK_\N\setminus\{P_0, \Pinf \}$.
Let  $\omega^{(3)}_{\Pinf,P_0}$ be a normal differential of
the third kind (cf.\ \eqref{a36} and \eqref{a37}) with
simple poles at $\Pinf$ and $P_0$ with residues $1$ and $-1$,
respectively. Thus
\begin{subequations} \lb{3.19d}
\begin{align}
\omega^{(3)}_{\Pinf,P_0} &
\underset{\zeta\to 0}{=}(\zeta^{-1}+\Oh(1))d\zeta
\text{ as $P\to\Pinf$}, \lb{3.19da} \\
\omega^{(3)}_{\Pinf,P_0} &
\underset{\zeta\to 0}{=}(-\zeta^{-1}+\Oh(1))d\zeta
\text{ as $P\to P_0$}, \lb{3.19db}
\end{align}
\end{subequations}
where
\begin{align}
&\zeta=\sigma/z^{1/2} \text{ for } P \text{ near }
\Pinf, \quad
\zeta=\sigma z^{1/2} \text{ for } P \text{ near } P_0,
\quad \sigma =\pm 1,
\lb{3.19dc} \\
& \hspace*{4cm} z^{1/2}=|z^{1/2}|e^{\frac{i}{2}\arg(z)}, \quad
0\leq\arg(z)<2\pi. \no
\end{align}
Hence we find
\begin{subequations} \lb{3.20k}
\begin{align}
\exp\bigg(-\int_{Q_0}^P \omega^{(3)}_{\Pinf,P_0}\bigg)
&\underset{\zeta\to 0}{=}e_{-1}\zeta^{-1}+e_0+\Oh(\zeta)
\text{ as $P\to\Pinf$}, \lb{3.20ka} \\
\exp\bigg(-\int_{Q_0}^P \omega^{(3)}_{\Pinf,P_0}\bigg)
& \underset{\zeta\to 0}{=}d_{1}\zeta + \Oh(\zeta^2)
\text{ as $P\to P_0$}, \lb{3.20kb}
\end{align}
\end{subequations}
for appropriate coefficients $e_{-1}$, $e_0$, and $d_{1}$
depending on the base point
$Q_{0}$. In the remainder of this paper we choose
$Q_{0}\in\calK_{n}\setminus\{P_{0},P_{\infty}\}$ such that
\begin{equation}
e_{-1}=1 \lb{3.20kc}
\end{equation}
in \eqref{3.20ka}. Let $\omega^{(2)}_{\Pinf,0}$ be
the normalized  differential of the second
kind holomorphic on $\calK_\N\setminus\{\Pinf\}$ such that
(cf.\ \eqref{a34}--\eqref{a35a})
\begin{equation}
\omega^{(2)}_{\Pinf,0}\underset{\zeta\to 0}{=}(\zeta^{-2}+
\Oh(1))d\zeta
\text{ as
$P\to\Pinf$} \lb{3.20b}
\end{equation}
and denote the vector of $b$-periods of
$\omega^{(2)}_{\Pinf,0}/(2\pi i)$
by ${\ul U}^{(2)}_0$ , that is,
\begin{equation}
{\ul U}^{(2)}_0 = (U^{(2)}_{0,1},\dots,U^{(2)}_{0,\N}),
\quad
U_{0,j}^{(0)}=\frac{1}{2\pi i} \int_{b_j}
\omega^{(2)}_{\Pinf,0} =-2c_j(n),
\quad j=1,\dots,\N, \lb{3.49ea}
\end{equation}
applying \eqref{a35a}.

\begin{theorem} \lb{theorem3.3}
Assume \eqref{3.14a}, \eqref{3.20kc}, $P\in\calK_\N
\setminus \{\Pinf\}$, and
$x, x_0\in\Omega$, where $\Omega\subseteq\bbR$ is an open
interval. Suppose that
$\calD_{\hat{\ul\mu}(x)}$, or equivalently,
$\calD_{\hat{\ul\nu}(x)}$
is nonspecial for $x\in\Omega$. Then $\phi(P,x)$ admits
the representation
\begin{align}
\begin{split}
\phi (P,x) = &\frac{\theta (\ul\Xi_{Q_0} - \ul A_{Q_0}
(\Pinf) + \ul\alpha_{Q_0} (\calD_{\hat{\ul\mu} (x)}))}
{\theta(\ul\Xi_{Q_0} -\ul A_{Q_0}
(\Pinf) +\ul\alpha_{Q_0} (\calD_{\hat{\ul\mu}(x)})+
\ul \Delta)} \times\\
& \times \frac{\theta (\ul\Xi_{Q_0} -
\ul A_{Q_0} (P) + \ul\alpha_{Q_0}
(\calD_{\hat{\ul\mu}(x)})+\ul \Delta)}{\theta (\ul\Xi_{Q_0}
-\ul A_{Q_0} (P) +
\ul\alpha_{Q_0}(\calD_{\hat{\ul\mu} (x)}))} \exp \bigg(
-\int_{Q_0}^P
\omega_{\Pinf,P_0}^{(3)}\bigg),
\lb{3.21}
\end{split}
\end{align}
with $\ul \Delta$ a halfperiod defined as
\begin{equation}
\ul \Delta = \ul A_{P_0}(\Pinf), \quad 2\ul \Delta=0
\pmod{L_\N}. \lb{3.24k}
\end{equation}
The components $\psi_j(P,x,x_0)$,
$j=1,2$ of the Baker--Akhiezer function $\Psi(P,x,x_0)$ read,
\begin{align}
\psi_1(P,x,x_0) = & \frac{\theta (\ul\Xi_{Q_0}
- \ul A_{Q_0} (P) + \ul\alpha_{Q_0}
(\calD_{\hat{\ul\mu}(x)})) \theta (\ul\Xi_{Q_0} -
\ul A_{Q_0} (\Pinf) + \ul\alpha_{Q_0}
(\calD_{\hat{\ul\mu}(x_0)}))}{ \theta (\ul\Xi_{Q_0} -
\ul A_{Q_0} (\Pinf) + \ul\alpha_{Q_0} (
\calD_{\hat{\ul\mu}(x)})) \theta (\ul\Xi_{Q_0} -
\ul A_{Q_0} (P) + \ul\alpha_{Q_0}
(\calD_{\hat{\ul\mu}(x_0)}))}
\times \no \\
& \times \exp \bigg(-i(x-x_0) \int_{Q_0}^P
\omega_{\Pinf,0}^{(2)}\bigg)
\lb{3.22}
\end{align}
and
\begin{align}
& \psi_2(P,x,x_0)  \no \\
& = -\frac{\theta (\ul\Xi_{Q_0}
- \ul A_{Q_0} (P) + \ul\alpha_{Q_0}(\calD_{\hat{\ul\mu}(x)})+
\ul \Delta)}
{\theta (\ul\Xi_{Q_0} -\ul A_{Q_0} (\Pinf) +
\ul\alpha_{Q_0} (\calD_{\hat{\ul\mu}(x)})+\ul \Delta)}
\frac{\theta (\ul\Xi_{Q_0} -\ul A_{Q_0} (\Pinf)
+ \ul\alpha_{Q_0}
(\calD_{\hat{\ul\mu}(x_0)}))}{\theta (\ul\Xi_{Q_0} -
\ul A_{Q_0} (P) + \ul\alpha_{Q_0} (\calD_{\hat{\ul\mu}(x_0)}))}
\times  \no \\
& \quad \times \exp \bigg(-i(x-x_0) \int_{Q_0}^P
\omega_{\Pinf,0}^{(2)}-\int_{Q_0}^P
\omega_{\Pinf,P_0}^{(3)}\bigg).
\lb{3.22a}
\end{align}
The Abel map linearizes the auxiliary divisors in
the sense that
\begin{subequations} \lb{3.23a}
\begin{align}
 \ul\alpha_{Q_0} (\calD_{\hat{\ul\mu}(x)}) &=
\ul\alpha_{Q_0} (\calD_{\hat{\ul\mu}(x_0)}) -
2i\ul c(\N)(x-x_0),
\lb{3.23aa} \\
 \ul\alpha_{Q_0} (\calD_{\hat{\ul\nu}(x)}) &=
\ul\alpha_{Q_0} (\calD_{\hat{\ul\nu}(x_0)}) -
2i\ul c(\N)(x-x_0),
\lb{3.23ab}
\end{align}
\end{subequations}
where $\ul c(\N)$ is defined in \eqref{A.7}.
The stationary solution $u(x)$ of $\sG_\N(u)=0$ finally reads
\begin{align}
u(x)&=u(x_0) \no\\
&\quad +2i\ln\left(
\f{\theta (\Pinf,\hat{\ul\mu}(x),\ul \Delta)
\theta (\Pinf,\hat{\ul\mu}(x_0))}
{\theta (\Pinf,\hat{\ul\mu}(x_0),\ul \Delta)
\theta (\Pinf,\hat{\ul\mu}(x)) }
\exp \big(- i e_0(x-x_0)\big) \right), \lb{3.24}
\end{align}
abbreviating
\begin{align}
\theta(P,\ul Q) & =\theta(\ul\Xi_{Q_0}-\ul A_{Q_0}(P)
+\ul\alpha_{Q_0}(\calD_{\ul Q})), \no \\
\theta(P,\ul Q, \ul \Delta) & =\theta(\ul\Xi_{Q_0}-\ul A_{Q_0}(P)
+\ul\alpha_{Q_0}(\calD_{\ul Q}) + \ul \Delta), \lb{3.24d} \\
& \hspace{4cm} \ul Q=(Q_1,\dots, Q_\N) \in \sigma^\N \calK_\N. \no
\end{align}
\end{theorem}
\begin{proof} Assume temporarily that
\begin{equation}
\mu_j(x)\neq \mu_{j'}(x), \, \nu_m(x)\neq \nu_{m'}(x)
\text{ for }
j\neq j', \, m\neq m', \text{ and }
 x\in\ti \Omega\subseteq\Omega,
\lb{3.24h}
\end{equation}
where $\ti \Omega$ is an open interval.
Denote the right-hand side of \eqref{3.21} by
$\Phi$. By the Riemann-Roch theorem \eqref{a42}, to
prove that
$\phi=\Phi$
it suffices to show that $\phi$ and $\Phi$ have the
same poles
and zeros as
well as the same value at one
point on $\calK_\N$.  From the definition  \eqref{3.4} of
$\phi$  we
conclude that it has simple zeros at
$\hat{\ul\nu}(x)$ and $P_0$ and simple poles at
$\hat{\ul\mu}(x)$
and $\Pinf$.
Note the linear equivalence $\calD_{\Pinf\hat{\ul\mu}(x)}\sim
\calD_{P_0\hat{\ul\nu}(x)}$,  that is,
\begin{equation}
\ul A_{Q_0} (\Pinf) + \ul\alpha_{Q_0} (\calD_{\hat {\ul\mu}(x)})
= \ul A_{Q_0} (P_0) + \ul\alpha_{Q_0}(\calD_{\hat {\ul\nu} (x)})
\lb{3.23c}
\end{equation}
and hence
\begin{equation}
\ul\alpha_{Q_0}(\calD_{\hat{\ul\nu}(x)})=
\ul\alpha_{Q_0}(\calD_{\hat{\ul\mu}(x)})+ \ul \Delta,
\lb{3.31}
\end{equation}
with
\begin{equation}
\ul \Delta = \ul A_{P_0}(\Pinf). \lb{3.31d}
\end{equation}
Since $P_0$ and $\Pinf$ are branch points of $\calK_\N$, the
right-hand side of \eqref{3.31d} is a half-period,
proving \eqref{3.24k}. By Theorem \ref{taa17a} one
concludes that $\Phi$ and $\phi$ share
the same zeros and poles. Thus $\Phi/\phi$ equals a
constant. From \eqref{3.4} one infers that
\begin{equation}
\phi(P,x)\underset{\zeta\to 0}{=}\zeta^{-1} +\Oh(1)
\text{ as $P\to\Pinf$},
\lb{3.25}
\end{equation}
and using the leading term in the expansion \eqref{3.20ka}
with $e_{-1}=1$ according to \eqref{3.20kc} then proves
$\phi=\Phi$ subject to \eqref{3.24h}.

Next we turn to the proof of \eqref{3.22}. One infers
from \eqref{3.9c}
that $\psi_1$ has first-order zeros at $\hat{\ul\mu}(x)$,
and first-order poles at
$\hat{\ul\mu}(x_0)$. Thus, by Theorem \ref{taa17a}, $\psi_1$
contains a factor
$\theta(P,\hat{\ul\mu}(x))/\theta(P,\hat{\ul\mu}(x_0))$.
A high-energy expansion (cf.\
\eqref{high}) then reveals that
\begin{equation}
\int_{x_0}^x dx'\, \f{iy(P)}{F_\N(z,x')}
\underset{\zeta\to 0}{=}
i(x-x_0)\zeta^{-1}+\Oh(1) \text{ as $P\to\Pinf$}, \lb{asy}
\end{equation}
and upon inserting \eqref{3.4} into \eqref{3.6a} one
concludes that $\psi_1$ has an essential singularity at
$\Pinf$ precisely of the type \eqref{asy}. Since
\begin{equation}
\int_{P_0}^P \omega_{\Pinf,0}^{(2)}\underset{\zeta\to 0}
{=}-\zeta^{-1}+\Oh(1) \text{ as
$P\to\Pinf$},
\end{equation}
the Riemann-Roch-type  uniqueness result, Lemma \ref{lem34},
applied to $\psi_1$ then yields \eqref{3.22} subject to
\eqref{3.24h}. The corresponding expression
\eqref{3.22a} for $\psi_2$ is then obvious from \eqref{3.6b},
\eqref{3.21}, and \eqref{3.22}.

Next we prove the linearization property \eqref{3.23aa} using
Lemma \ref{lemma3.2} (still
assuming \eqref{3.24h}). Equation \eqref{3.23ab} then
follows from
\eqref{3.23aa} and \eqref{3.23c}.
 From
\begin{equation}
\ul\alpha_{Q_0}(\calD_{\hat{\ul\mu}(x)})=
\bigg(\sum_{j=1}^\N \int_{Q_0}^{\hat\mu_j(x)}\ul\omega\bigg)
\pmod{L_\N}
\lb{3.27}
\end{equation}
and the Dubrovin equations \eqref{3.15} one infers
\begin{align}
\f{d}{dx} \ul\alpha_{Q_0}(\calD_{\hat{\ul\mu}(x)})&=
\sum_{j=1}^\N \mu_j^\prime(x)\sum_{k=1}^\N\ul
c(k)\f{\mu_j(x)^{k-1}}{y(\hat\mu_j(x))} \no\\
&=-2i\sum_{j,k=1}^\N \ul c(k) \f{\mu_j(x)^{k-1}}
{\prod^\N_{\substack{\ell=1\\ \ell\neq
j}}(\mu_j(x)-\mu_\ell(x))}, \quad x\in\ti \Omega.\lb{3.28}
\end{align}
Lagrange's interpolation formula (cf.\ Theorem \ref{theoremB1})
\begin{equation}
\sum_{j=1}^\N\mu_j^{k-1}
\prod_{\substack{\ell=1 \\ \ell \neq j}}^{\N}
(\mu_{j} - \mu_{\ell})^{-1}=
\delta_{k,\N},\quad k=1,\dots,\N \label{3.29}
\end{equation}
then yields
\begin{equation}
\f{d}{dx} \ul\alpha_{Q_0}(\calD_{\hat{\ul\mu}(x)})=-2 i \ul c(\N)
=i {\ul U}_0, \quad x\in\ti \Omega
\label{3.30}
\end{equation}
and hence \eqref{3.23aa} subject to \eqref{3.24h}. Moreover,
applying \eqref{a35a} establishes the relation between the
left-hand side of
\eqref{3.30}  and the vector ${\ul U}_0$ of $b$-periods of
$\omega^{(2)}_{\Pinf,0}/(2\pi i)$ introduced in
\eqref{3.49ea}.

To prove formula \eqref{3.24} for $u(x)$ we employ a
high-energy expansion
for $\phi(P,x)$  and the Riccati equation \eqref{3.8a}.
First we
conclude from \eqref{b27}
\begin{equation}
\ul\omega\underset{\zeta\to 0}{=}(-2\ul c(\N)+
\Oh(\zeta^2))d\zeta \text{ as
$P\to
P_\infty$,}\lb{1.3.62g}
\end{equation}
and hence
\begin{align}
\ul A_{Q_0} (P)&=\int_{Q_0}^P \ul\omega \pmod{L_\N}
\underset{\zeta\to 0}{=}\ul A_{Q_0}(P_\infty)-2\ul
c(\N)\zeta+\Oh(\zeta^3) \no \\
&=\ul A_{Q_0}(P_\infty)+\ul U_0^{(2)}\zeta+\Oh(\zeta^2),
\lb{1.3.62h}
\end{align}
using \eqref{a35a} and \eqref{3.49ea}. Expanding the
subsequent ratios
of Riemann theta functions in \eqref{3.21} one finds
\begin{align}
&\f{\theta(P,\hat{\ul\mu}(x))}
{\theta(\Pinf,\hat{\ul\mu}(x))} \no \\
&\underset{\zeta\to 0}{=}
1-\f{\sum_{j=1}^\N U_{0,j}^{(2)}\f{\partial}{\partial w_j}
\theta (\ul {\Xi}_{Q_0}-\ul A_{Q_0}(P_\infty)+\ul w+
\ul \alpha_{Q_0}(\hat {\ul \mu}(x)))\vert_{\ul w=0}}
{\theta (\ul \Xi_{Q_0}-\ul A_{Q_0}(P_\infty)+
\ul \alpha_{Q_0}(\hat {\ul \mu}(x)))}\zeta
+\Oh(\zeta^3),\lb{1.3.62i}
\end{align}
and the same formula for the theta function ratio involving the
additional half-period $\ul \Delta$. Here $\sum_{j=1}^\N
U_{0,j}^{(2)}\partial/\partial w_j$
denotes the directional derivative in $\ul U_{0}^{(2)}$-direction.
Given \eqref{3.49ea} we may write
\begin{align}
&\sum_{j=1}^\N i U_{0,j}^{(2)}\f{\partial}{\partial w_j}
\theta (\ul \Xi_{Q_0}-\ul A_{Q_0}(P_\infty)+\ul w+
\ul \alpha_{Q_0}(\hat {\ul \mu}(x_0))+i\ul
U_{0}^{(2)}(x-x_0))\vert_{\ul w =0} \no \\
&\quad = \f{d}{dx}\theta (\ul \Xi_{Q_0}-
\ul A_{Q_0}(P_\infty)+\ul \alpha_{Q_0}(\hat {\ul \mu}(x_0))+i\ul
U_{0}^{(2)}(x-x_0)), \label{1.3.63a}
\end{align}
and hence obtain from \eqref{1.3.62i},
\begin{equation}
\f{\theta(P,\hat{\ul\mu}(x))}{\theta(\Pinf,\hat{\ul\mu}(x))}
\underset{\zeta\to 0}{=}
1+i\f{d}{dx}\ln(\theta(\Pinf,\hat{\ul\mu}(x)))\zeta
+\Oh(\zeta^3) \text{ as } P \to \Pinf, \lb{3.27a}
\end{equation}
and the identical formula for the theta function ratio
involving $\ul \Delta$. Together with \eqref{3.20ka} this shows that
\begin{equation}
\phi(P,x)\underset{\zeta\to 0}{=}\zeta^{-1}+
i\f{d}{dx}\ln\bigg(\f{\theta(\Pinf,\hat{\ul\mu}(x),
\ul \Delta)}
{\theta(\Pinf,\hat{\ul\mu}(x))}\bigg)
+e_0+\Oh(\zeta) \text{ as } P \to \Pinf. \lb{3.28b}
\end{equation}
Expanding the Riccati equation \eqref{3.8a} for $P$ near
$\Pinf$ then yields
\begin{equation}
u_x=2i\f{d}{dx}\ln\bigg(\f{\theta(\Pinf,\hat{\ul\mu}(x),
\ul \Delta)}
{\theta(\Pinf,\hat{\ul\mu}(x))}\bigg)
+2e_0, \quad x\in\ti \Omega \lb{3.29b}
\end{equation}
and hence \eqref{3.24}, assuming \eqref{3.24h}. The
extension of all
these results from $x\in\ti \Omega$ to $x\in\Omega$ then simply
follows from the continuity of $\ul \alpha_{P_0}$ and
the hypothesis of $\calD_{\ul {\hat \mu}(x)}$ being
nonspecial for $x\in\Omega$.

To verify that
$u(x)$ indeed satisfies
$\N$th stationary sGmKdV equation it suffices to note
\begin{equation}
(V_{\N,x}-[U,V_\N])\Psi=\f{d}{dx}(V_\N\Psi+\f{y}{z}\Psi)=0,
\end{equation}
applying \eqref{3.9a} and \eqref{3.9ab} repeatedly. Since
$P\in\calK_\N\setminus \{\Pinf\}$ is arbitrary, one concludes
$V_{\N,x}-[U,V_\N]=0$ which completes the proof.
\end{proof}

While this approach to the algebro-geometric solutions of
the sGmKdV hierarchy resembles that of the AKNS hierarchy
(which includes the mKdV hierarchy) in many ways
(cf.\ \cite{GR96}), there
are, however, some characteristic differences. In particular, the branch
point
$P_0=(0,0)$ is an unusual
necessity in the sG context and $\Pinf$ (as in the KdV context)
is a branch point as opposed to the AKNS case. This shows
that sGmKdV
curves are actually special KdV curves (with $0$ a branch
point).

\section{The time-dependent sGmKdV formalism} \lb{s4}

In our final section we indicate how to employ the
polynomial formalism of Sections \ref{s2} and \ref{s3} to
derive the algebro-geometric solutions of the sGmKdV
hierarchy.

The basic problem in the analysis of algebro-geometric solutions
of completely
integrable soliton equations is to solve the time-dependent
$r$th sGmKdV equation
with
initial data a stationary
solution of the $\N$th equation in the hierarchy.
More precisely, consider a solution
$u^{(0)}(x)$ of the $\N$th stationary sGmKdV equation,
that is,
\begin{equation}
g_{\N-1,x}-i(\beta e^{iu^{(0)}}-\alpha e^{-iu^{(0)}})=0,
\quad \N\in\bbN, \lb{4.1}
\end{equation}
subject to the constraint
\begin{equation}
\alpha \beta = P_{2\N}(0). \lb{4.1a}
\end{equation}
Let $r\in\bbN_0$ be fixed (and independent of $\N$).  We want
to find a solution $u$ of
$\sG_r(u)=0$ with $u(x,t_{0,r})=u^{(0)}(x)$.
To emphasize that the integration constants that enter in
the definitions of the sGmKdV
hierarchy may be different for the stationary initial data
and the time-dependent equation,
we indicate this by adding a tilde on all the time-dependent
quantities. Hence we shall employ the notation $\ti V_r$,
$\ti F_r$, $\ti G_{r-1}$, $\ti H_r$, $\tilde f_{\ell}$,
$\tilde g_{\ell}$, $\tilde h_{\ell}$, $\tilde c_{\ell}$,
$\tilde \alpha$,
$\tilde \beta$ in
order to distinguish them from  $V_\N$,
$F_\N$, $G_{\N-1}$, $H_\N$, $f_{\ell}$,
$g_{\ell}$, $h_{\ell}$, $c_{\ell}$, $\alpha$, $\beta$,
in the following.

Thus we are seeking a solution $u$ of
\begin{equation}
\sG_r(u)=u_{xt_r}+2i\tilde g_{r-1,x}-i(\tilde\beta e^{iu}-
\tilde\alpha e^{-iu})=0,
\quad u(x,t_{0,r})=u^{(0)}(x). \lb{4.2}
\end{equation}
Actually, since we need to rely on the isospectral
property of the sGmKdV flows we go a step further and assume
 \eqref{4.1} not only at $t_r=t_{0,r}$ but for all $t_r\in\bbR$.
Together with \eqref{4.2} we are thus led to start with
\begin{equation}
U_{t_r}-\ti V_{r,x}+[U,\ti V_r]=0, \quad V_{\N,x}=[U,V_\N],
\quad (x,t_r)\in\bbR^2. \lb{4.3}
\end{equation}
Explicitly, this yields for each entry of $V_\N$ and
$\ti V_r$ (cf.\ \eqref{2.4}, \eqref{2.34})
\begin{subequations} \lb{4.4}
\begin{align}
F_{\N,x}&=-i u_x F_\N-2iz G_{\N-1},  \lb{4.4a}\\
H_{\N,x}&=i u_x H_\N+2iz G_{\N-1},  \lb{4.4b}\\
G_{\N-1,x}&=i(H_\N-F_\N). \lb{4.4c}
\end{align}
\end{subequations}
and
\begin{subequations}\lb{4.6}
\begin{align}
u_{xt_r}&=-2i\ti G_{r-1,x}-2(\ti H_r- \ti F_r), \lb{4.6a} \\
\ti F_{r,x} &=-i u_x \ti F_r-2iz\ti G_{r-1},  \lb{4.6b} \\
\ti H_{r,x}&=i u_x \ti H_r+2iz \ti G_{r-1}.  \lb{4.6c}
\end{align}
\end{subequations}

At this point we proceed assuming there is a solution of
\eqref{4.2}, or
rather, equations \eqref{4.4} and
\eqref{4.6}. Once explicit expressions for algebro-geometric
solutions are obtained, it is a straightforward manner to
verify directly that they satisfy \eqref{4.3} (resp.,
\eqref{4.4} and \eqref{4.6}).

It will turn out later (cf.\ Remark \ref{remark4.3}) that
$\alpha,\beta,\tilde \alpha,\tilde \beta$ in \eqref{4.1} and
\eqref{4.2} are not independent of each other but constrained
by
\begin{equation}
f_\N \tilde h_r=\tilde f_r h_\N \text{ or } \alpha\tilde \beta =
\tilde \alpha \beta.
\lb{4.6d}
\end{equation}

Throughout this section we assume \eqref{4.3}
(resp.\ \eqref{4.4} and
\eqref{4.6}) and in accordance with
Section \ref{s2} we suppose that
$F_\N$,
$G_{\N-1}$, $H_\N$, $\ti F_r$, $\ti G_{r-1}$, and $\ti H_r$
have the polynomial structure \eqref{2.14}, \eqref{3.7}, with
$t_r\in\bbR$ the additional deformation
parameter in $u$. We also recall our conventions in
\eqref{2.22} and
\eqref{2.28k} and set
\begin{equation}
\ti G_{-1}(x,t_0)=0, \quad \tilde g_{-1}(x,t_0)=0
\text{ for } r=0. \lb{4.6e}
\end{equation}
Hence \eqref{2.3}--\eqref{2.20} apply to $F_\N$, $G_{\N-1}$,
$H_\N$, $f_j$, $g_j$, and $h_j$ and \eqref{2.14},
\eqref{2.16}--\eqref{2.27},
\eqref{2.28} for $j=0,\dots,\N-2$,
\eqref{2.17aa}--\eqref{2.36}, with $\N\to r$,
$c_{\ell}\to\tilde c_{\ell}$,
$\alpha \to \tilde \alpha$, $\beta \to \tilde \beta$
apply to $\ti F_r$,
$\ti G_{r-1}$, $\ti H_r$, $\tilde f_j$, $\tilde g_j$,
and $\tilde h_j$.
In particular, the fundamental
identity \eqref{2.7} holds,
\begin{equation}
z^2 G_{\N-1}(z,x,t_r)^2+zF_\N(z,x,t_r) H_\N(z,x,t_r)=
R_{2\N+1}(z), \lb{4.5}
\end{equation}
with
\begin{equation}
F_\N(z,x,t_r)=\prod_{j=1}^\N(z-\mu_j(x,t_r)), \quad
H_\N(z,x, t_r)=\prod_{j=1}^\N(z-\nu_j(x,t_r)), \lb{4.8}
\end{equation}
assuming \eqref{3.1a} for the remainder of this section.
Hence the hyperelliptic curve
$\calK_\N$ is still given by
\begin{equation}
\calK_\N \colon \calF_\N(z,y)=y^2-R_{2\N+1}(z)=0. \lb{4.7}
\end{equation}
In analogy to equations \eqref{3.12} and \eqref{3.4} we define
\begin{subequations} \lb{4.10}
\begin{align}
\hat\mu_j(x,t_r)&=(\mu_j(x,t_r),-\mu_j(x,t_r)
G_{\N-1}(\mu_j(x,t_r),x,t_r))\in\calK_\N,
\quad j=1,\dots,\N, \lb{4.10a}\\
\hat\nu_j(x,t_r)&=(\nu_j(x,t_r),\nu_j(x,t_r)
G_{\N-1}(\nu_j(x,t_r),x,t_r))\in\calK_\N, \quad
j=1,\dots,\N.\lb{4.10b}
\end{align}
\end{subequations}
and
\begin{align}
&\phi(P,x,t_r)=\f{y(P)-zG_{\N-1}(z,x,t_r)}{F_\N(z,x,t_r)}
=\f{zH_{\N}(z,x,t_r)}{y(P)+zG_{\N-1}(z,x,t_r)}, \lb{4.9} \\
& \hspace*{4cm} (x,t_r)\in\bbR^2, \,
P=(z,y)\in\calK_\N\setminus\{\Pinf \} \no
\end{align}
Hence the divisor $(\phi(P,x,t_r))$ of
$\phi(P,x,t_r)$ reads
\begin{equation}
(\phi(P,x,t_r))=\calD_{P_0\hat{\ul\nu}(x,t_r)}-
\calD_{\Pinf\hat{\ul\mu}(x,t_r)},
\lb{4.10d}
\end{equation}
with
\begin{equation}
\hat{\ul\mu}(x,t_r)=(\hat\mu_1(x,t_r),\dots,\hat\mu_\N(x,t_r)),
\quad
\hat{\ul\nu}(x,t_r)=(\hat\nu_1(x,t_r),\dots,\hat\nu_\N(x,t_r)).
\lb{3.10e}
\end{equation}
The time-dependent Baker--Akhiezer function
\begin{equation}
\Psi(P,x,x_0,t_r,t_{0,r})=
\begin{pmatrix} \psi_1(P,x,x_0,t_r,t_{0,r}) \\
\psi_2(P,x,x_0,t_r,t_{0,r})
\end{pmatrix}\lb{4.11}
\end{equation}
is defined by
\begin{align}
&\psi_1(P,x,x_0,t_r,t_{0,r})=
\exp\bigg(-\f{i}2(u(x,t_r)-u(x_0,t_{r}))
\lb{4.12a}\\
&\qquad +i\int_{x_0}^x dx'\,\phi(P,x',t_r)-
\int_{t_{0,r}}^{t_r}ds\,\big(\f1z \ti F_r(z,x_0,s)\phi(P,x_0,s)+
\ti G_{r-1}(z,x_0,s) \big)\bigg),\no \\
&\psi_2(P,x,x_0,t_r,t_{0,r})=
-\psi_1(P,x,x_0,t_r,t_{0,r}) \phi(P,x,t_r),
\lb{4.12b} \\
&\hspace*{6cm} P\in\calK_\N\setminus\{\Pinf\}, \,
(x,x_0,t_r,t_{0,r})\in\bbR^4. \no
\end{align}
The properties of $\phi$ can now be summarized as follows.
\begin{lemma} \lb{lemma4.1}
Assume \eqref{3.1a}, $P=(z,y)\in\calK_\N\setminus\{\Pinf \}$,
and let
$(z,x,x_0,t_r,t_{0,r})\in\bbC\times\bbR^4$. Then the function
$\phi$ satisfies
\begin{align}
&\phi(P,x,t_r)^2-i\phi_x(P,x,t_r)=z+u_x(x,t_r)\phi(P,x,t_r),
\lb{4.13a}\\
&\phi_{t_r}(P,x,t_r)=\ti F_r(z,x,t_r)-\ti H_r(z,x,t_r)+
\f{i}{z}(\phi(P,x,t_r)\ti F_r(z,x,t_r))_x, \lb{4.13b} \\
&\phi(P,x,t_r)+\phi(P^*,x,t_r)=-2z\f{G_{\N-1}(z,x,t_r)}
{F_\N(z,x,t_r)},
\lb{4.13c}\\
&\phi(P,x,t_r)\phi(P^*,x,t_r)=-z\f{H_{\N}(z,x,t_r)}
{F_\N(z,x,t_r)},
\lb{4.13d}\\
&\phi(P,x,t_r)-\phi(P^*,x,t_r)=2\f{y(P)}{F_\N(z,x,t_r)}.
\lb{4.13e}
\end{align}
\end{lemma}
\begin{proof}
Equation \eqref{4.13a} follows as in Lemma \ref{lemma3.1}.
To prove
\eqref{4.13b} we first
observe that
\begin{equation}
\big(\partial_x+i(2\phi-u_x)\big)(\phi_{t_r}+
(\ti H_r - \ti F_r) -\f{i}{z}(\phi\ti F_r)_x)=0 \lb{4.15}
\end{equation}
using \eqref{4.13a} and relations \eqref{4.6} repeatedly.
Thus,
\begin{equation}
\phi_{t_r}+(\ti H_r - \ti F_r)- \f{i}{z}(\phi\ti F_r)_x
=C\exp\bigg(-i\int^x dx' \, (2\phi-u_{x'})\bigg), \lb{4.16}
\end{equation}
where the left-hand side is meromorphic at $\Pinf$, while
the right-hand
side is meromorphic
at $\Pinf$ only if $C=0$, which proves \eqref{4.13b}.
Equations
\eqref{4.13c}--\eqref{4.13e}
are proved as in Lemma \ref{lemma3.1}.
\end{proof}
Next we prove that relations \eqref{4.4} and \eqref{4.6}
determine the time-development of $F_\N$, $G_\N$, and $H_\N$.

\begin{lemma} \label{lemma4.2}
Suppose \eqref{3.1a} and let
$(z,x,x_0,t_r,t_{0,r})\in\bbC\times\bbR^4$.  Then
\begin{subequations}\lb{4.18}
\begin{align}
F_{\N,t_r}&=2(G_{\N-1}\ti F_r-\ti G_{r-1} F_\N), \lb{4.18a} \\
zG_{\N-1,t_r}&=(F_\N \ti H_r-\ti F_r H_\N), \lb{4.18b} \\
H_{\N,t_r}&=2(\ti G_{r-1} H_\N- G_{\N-1} \ti H_r). \lb{4.18c}
\end{align}
\end{subequations}
Equations \eqref{4.18} are equivalent to\footnote{Equation
\eqref{4.19} is almost equivalent  with the consistency
requirement that
$V_{\N,x t_r}=V_{\N,t_r x}$. Using
\eqref{4.3} we find $[\tilde V_r,V_\N]_x=V_{\N,t_r x}$. This
shows that
$[\tilde V_r,V_\N]$ equals $V_{\N,t_r}$ up to a $2 \times 2$
matrix
depending on $z$ and $t_r$, but not on $x$.
\eqref{4.19} shows that this matrix is identically zero.}
\begin{equation}
V_{\N,t_r}=[\ti V_r,V_\N].\lb{4.19}
\end{equation}
\end{lemma}
\begin{proof}
We prove \eqref{4.18a} by using \eqref{4.13e} which shows that
\begin{equation}
(\phi(P)-\phi(P^*))_{t_r}=-2y(P)\f{F_{\N,t_r}}{F_\N^2}.
\lb{4.20}
\end{equation}
However, the left-hand side of \eqref{4.20} also equals
\begin{equation}
\phi(P)_{t_r}-\phi(P^*)_{t_r}=\f{4y(P)}
{F_\N^2}(\ti G_{r-1} F_\N-\ti F_r
G_{\N-1}), \lb{4.21}
\end{equation}
using \eqref{4.13b}, \eqref{4.13e}, \eqref{4.4a}, and
\eqref{4.6a}.
Combining \eqref{4.20} and
\eqref{4.21} proves \eqref{4.18a}.  Similarly, to prove
\eqref{4.18b}, we
use \eqref{4.13c} to write
\begin{equation}
(\phi(P)+\phi(P^*))_{t_r}=-
\f{2z}{F_\N^2}\big(G_{\N-1,t_r}F_\N-G_{\N-1}F_{\N,t_r
}\big). \lb{4.22}
\end{equation}
Here  the left-hand side can be expressed as
\begin{equation}
\phi(P)_{t_r}+\phi(P^*)_{t_r}=-2\ti H_r+2\f{H_\N}{F_\N}\ti
F_r+4z(\f{G_{\N-1}}{F_\N})^2 \ti F_r-
4z\f{G_{\N-1}}{F_\N} \ti G_{r-1}, \lb{4.23}
\end{equation}
using \eqref{4.4a} and \eqref{4.6a}.  Combining
\eqref{4.22} and
\eqref{4.23}, using \eqref{4.18a},
proves \eqref{4.18b}.  Finally, \eqref{4.18c} follows
by differentiating
\eqref{4.5}, that is,
$(z G_{\N-1})^2+z F_\N H_\N=R_{2\N+1}$, with respect to
$t_r$, and using
\eqref{4.18a} and
\eqref{4.18b}.
\end{proof}

\begin{remark} \lb{remark4.3}
Taking $z=0$ in \eqref{4.18b}, one infers the compatibility
relation
\begin{equation}
f_\N(x) \tilde h_r(x)=\tilde f_r(x) h_\N(x), \lb{4.30a}
\end{equation}
or equivalently, using $f_\N=\alpha e^{-iu}$,
$\tilde f_r=\tilde \alpha e^{-iu}$,
$h_\N=\beta e^{iu}$, and $\tilde h_r=\tilde \beta e^{iu}$,
one obtains
the constraint
\begin{equation}
\alpha \tilde \beta=\tilde \alpha \beta. \lb{4.19d}
\end{equation}
\end{remark}

Lemmas \ref{lemma4.1} and \ref{lemma4.2} permit us to
characterize $\Psi$.

\begin{lemma} \lb{lemma4.4}
Assume \eqref{3.1a}, $P=(z,y)\in\calK_\N\setminus\{\Pinf \}$,
and let
$(x,x_0,t_r,t_{0,r})\in\bbR^4$. Then the Baker--Akhiezer
function $\Psi$ is meromorphic on
$\calK_\N\setminus\{\Pinf\}$
and satisfies
\begin{align}
&\Psi_x(P,x,x_0,t_r,t_{0,r})=
U(z,x,t_r)\Psi(P,x,x_0,t_r,t_{0,r}), \lb{4.14a} \\
&-\f{y(P)}{z}\Psi(P,x,x_0,t_r,t_{0,r})=
V_\N(z,x,t_r)\Psi(P,x,x_0,t_r,t_{0,r}),
\lb{4.14ab} \\
&\Psi_{t_r}(P,x,x_0,t_r,t_{0,r})=
\ti V_r(z,x,t_r)\Psi(P,x,x_0,t_r,t_{0,r}),
\lb{4.14b} \\
&\psi_1(P,x,x_0,t_r,t_{0,r}) \lb{4.24a}\\
&=\bigg(\f{F_\N(z,x,t_r)}{F_\N(z,x_0,t_{0,r})}\bigg)^{1/2}
\exp\bigg(iy(P)\int_{x_0}^x \f{dx'}{F_\N(z,x',t_r)}
 -\f{y(P)}{z}\int_{t_{0,r}}^{t_r}ds\, \f{\ti
F_r(z,x_0,s)}{F_\N(z,x_0,s)}\bigg), \no \\
&\psi_1(P,x,x_0,t_r,t_{0,r})\psi_1(P^*,x,x_0,t_r,t_{0,r})
=\f{F_\N(z,x,t_r)}{F_\N(z,x_0,t_{0,r})},\lb{4.24c} \\
&\psi_2(P,x,x_0,t_r,t_{0,r})\psi_2(P^*,x,x_0,t_r,t_{0,r})
=-z\f{H_\N(z,x,t_r)}{F_\N(z,x_0,t_{0,r})},\lb{4.24d} \\
&\psi_1(P,x,x_0,t_r,t_{0,r})\psi_2(P^*,x,x_0,t_r,t_{0,r})\no \\
&\qquad\qquad+\psi_1(P^*,x,x_0,t_r,t_{0,r})
\psi_2(P,x,x_0,t_r,t_{0,r})
=2z\f{G_{\N-1}(z,x,t_r)}{F_\N(z,x_0,t_{0,r})}, \lb{4.24e} \\
&\psi_1(P,x,x_0,t_r,t_{0,r})\psi_2(P^*,x,x_0,t_r,t_{0,r})\no \\
&\qquad\qquad-\psi_1(P^*,x,x_0,t_r,t_{0,r})
\psi_2(P,x,x_0,t_r,t_{0,r})
=2\f{y(P)}{F_\N(z,x_0,t_{0,r})}. \lb{4.24f}
\end{align}
\end{lemma}
\begin{proof}
Relations \eqref{4.14a} and \eqref{4.14ab} follow as in Lemma
\ref{lemma3.1}, while the time
evolution
\eqref{4.14b} is a consequence of the definition of $\Psi$
in \eqref{4.12a}, \eqref{4.12b} as well as \eqref{4.13b},
rewriting
\begin{equation}
\phi_{t_r}=\big(\f{i}z  \phi\ti F_r+i \ti G_{r-1}+
\f12 u_{t_r}\big)_x,
\lb{4.17}
\end{equation}
using \eqref{4.6a} and
\begin{equation}
\phi_{t_r}=-\ti H_r+\f{1}z \ti F_r \phi^2+2\phi\ti G_{r-1},
\lb{4.17a}
\end{equation}
using \eqref{4.13a} and \eqref{4.6b}. To prove \eqref{4.24a}
we recall the definition
\eqref{4.12a}, that is,
\begin{align}
&\psi_1(P,x,x_0,t_r,t_{0,r})=
\exp\bigg(-\f{i}2(u(x,t_r)-u(x_0,t_{r}))
\lb{4.25}\\
&\quad +i\int_{x_0}^x dx'\,\phi(P,x',t_r)-
\int_{t_{0,r}}^{t_r}ds\,(\f1z \ti F_r(z,x_0,s)\phi(P,x_0,s)+\ti
G_{r-1}(z,x_0,s)) \bigg), \no \\
&=\exp\bigg(iy(P)\int_{x_0}^x dx'\,
\f{1}{F_\N(z,x',t_r)}-\f{y(P)}{z}\int_{t_{0,r}}^{t_r}ds\,
\f{\ti F_r(z,x_0,s)}{F_\N(z,x_0,s)}\no \\
&\quad -\f{i}2(u(x,t_r)-u(x_0,t_{r}))-iz\int_{x_0}^x dx'\,
\f{G_{\N-1}(z, x', t_r)}{F_\N(z,x',t_r)} \no \\
&\quad +\int_{t_{0,r}}^{t_r}ds\,
\big(\f{\ti F_r(z,x_0,s)}{F_\N(z,x_0,s)}G_{\N-1}(z,x_0,s)-\ti
G_{r-1}(z,x_0,s)\big)\bigg), \no
\end{align}
using \eqref{4.9}.  By first invoking \eqref{4.4a} and
subsequently
\eqref{4.18a}, relation
\eqref{4.24a} follows. Evaluating \eqref{4.24a} at the points
$P$ and $P^*$
and multiplying
the resulting expressions yields \eqref{4.24c}. The remaining
statements
are direct
consequences of \eqref{4.13c}--\eqref{4.13e} and \eqref{4.24a}.
\end{proof}

We also note the following trace formulas, the $t_r$-dependent
analog
of \eqref{3.10}.
\begin{lemma}\lb{lemma4.5}
Suppose \eqref{3.1a}. Then $u(x,t_r)$ satisfies
\begin{subequations}\lb{4.33}
\begin{align}
u(x,t_r)&=i\ln\bigg((-1)^\N\alpha^{-1}
\prod_{j=1}^\N\mu_j(x,t_r) \bigg),
 \lb{4.33a}\\
u(x,t_r)&=-i\ln\bigg((-1)^\N\beta^{-1}
\prod_{j=1}^\N\nu_j(x,t_r) \bigg),
 \lb{4.33b}
\end{align}
\end{subequations}
where $\alpha\beta=\prod_{m=1}^{2\N}E_m \neq 0$.
\end{lemma}
\begin{proof}
The proof is identical to that of relations \eqref{3.10}.
\end{proof}
Next we turn to the time evolution of the quantities $\mu_j$
and $\nu_j$
and, as in Section \ref{s3}, we will assume that $\calK_\N$
is nonsingular
for the rest of this section.
\begin{lemma}\lb{lemma4.6}
Assume \eqref{3.14a} and suppose the zeros
$\{\mu_j(x,t_r) \}_{j=1,\dots,\N}$ of $F_\N(\dott,x,t_r)$ remain
distinct for $(x,t_r)\in\Omega$, where
$\Omega\subseteq\bbR^2$ is open and connected. Then
$\{\mu_j(x,t_r)\}_{j=1,\dots,\N}$ satisfies the following
system of differential equations
\begin{align}
\mu_{j,x}(x,t_r)&=-2i\f{y(\hat\mu_j(x,t_r))}
{\prod_{\substack{\ell=1\\ \ell\neq j}}^\N(\mu_j(x,t_r)-
\mu_\ell(x,t_r))},
\lb{4.26}\\
\mu_{j,t_r}(x,t_r)&=2\f{\ti F_r(\mu_j(x,t_r),x,t_r)}{\mu_j(x,t_r)}
\f{y(\hat\mu_j(x,t_r))}{\prod_{\substack{\ell=1\\ \ell\neq
j}}^\N(\mu_j(x,t_r)-\mu_\ell(x,t_r))},
\lb{4.26a} \\
& \hspace*{5cm} j=1, \dots, \N, \, (x,t_r)\in\Omega. \no
\end{align}
with initial conditions
\begin{equation}
\hat\mu_j(x_0,t_{0,r})\in\calK_\N,\quad j=1, \dots, \N, \lb{4.27}
\end{equation}
for some fixed $(x_0,t_{0,r})\in\Omega$.  The initial value
problem \eqref{4.26}--\eqref{4.27} has a unique
solution
$\{\hat\mu_j(x,t_r)\}_{j=1,\dots,\N}$ such that
\begin{equation}
\hat\mu_j(x,t_r)\in C^\infty(\Omega,\calK_\N),
\quad j=1, \dots, \N. \lb{4.28}
\end{equation}
For the zeros $\{\nu_j(x,t_r) \}_{j=1,\dots,\N}$ of
$H_\N(\dott,x,t_r)$
identical statements hold
with $\mu$ replaced by $\nu$. In particular,
$\{\hat\nu_j(x,t_r)\}_{j=1,\dots,\N}$ satisfies
\begin{align}
\nu_{j,x}(x,t_r)&=-2i\f{y(\hat\nu_j(x,t_r))}
{\prod_{\substack{\ell=1\\ \ell\neq j}}^\N(\nu_j(x,t_r)-
\nu_\ell(x,t_r))},
\lb{4.29} \\
\nu_{j,t_r}(x,t_r)&=2\f{\ti H_r(\nu_j(x,t_r),x,t_r)}{\nu_j(x,t_r)}
\f{y(\hat\nu_j(x,t_r))}{\prod_{\substack{\ell=1\\ \ell\neq
j}}^\N(\nu_j(x,t_r)-\nu_\ell(x,t_r))},
\lb{4.29a} \\
& \hspace*{5cm} j=1, \dots, \N, \, (x,t_r)\in\Omega. \no
\end{align}
\end{lemma}
\begin{proof}
It suffices to prove \eqref{4.26a} since the argument for
\eqref{4.29a}
is analogous and that for \eqref{4.26} and \eqref{4.29}
has been given in the proof of Lemma \ref{lemma3.2}.
Inserting
$z=\mu_j(x,t_r)$ into \eqref{4.18a}, observing
\eqref{4.10a}, yields
\begin{equation}
F_{\N,t_r}(\mu_j)=-\mu_{j,t_r}
\prod_{\substack{\ell=1\\ \ell\neq j}}^\N(\mu_j-\mu_{\ell})
=2\f{\ti F_r(\mu_j)}{\mu_j} \mu_j G_{\N-1}(\mu_j)
=-2 \f{\ti F_r(\mu_j)}{\mu_j} y(\hat \mu_j). \lb{4.30}
\end{equation}
The rest is identical to the proof of Lemma \ref{lemma3.2}.
\end{proof}

\begin{remark}\lb{remark4.7}
Consider the case $\N\in\bbN$, and $r=0$, and $\alpha=\beta=i/4$.
Then, as explicitly proven in \cite{AA87}, \cite{EF85},
\cite{FM82}, \cite{TCL87}, \cite{TTCL84},
\begin{equation}
u(x,t_0)=i\ln\bigg((-1)^\N \alpha^{-1} \prod_{j=1}^\N\mu_j(x,t_0)
\bigg)
\end{equation}
satisfies
\begin{equation}
u_{xt_0}=\sin(u)
\end{equation}
whenever $\{\hat \mu_j(x,t_0)\}_{j=1,\dots,\N}$
satisfies \eqref{4.26} and \eqref{4.26a} for $r=0$. As discussed
in detail in \cite{EGHL97}, the trace relation \eqref{4.33a} for
$u(x,t_r)$ subject to the Dubrovin-type equations \eqref{4.26},
\eqref{4.26a} yields algebro-geometric solutions of (higher-order)
sGmKdV equations. In fact, a systematic study of $\ti F_r(\mu_j)$ in
terms of certain symmetric functions of $\mu_1,\dots,\mu_\N$
(cf.\ Appendix \ref{B}) yields an alternative approach to the
sGmKdV hierarchy.
\end{remark}

Next we turn to the principal result of this section, the
representation of $\phi$, $\Psi$, and $u$
in terms of the Riemann theta function associated with
$\calK_\N$. First we need to introduce
some notation.  Let $\omega^{(2)}_{\Pinf,2q}$
be a normalized differential of second kind with unique pole
at $\Pinf$ and
principal part
$\zeta^{-2q-2}d\zeta$ near $\Pinf$.  Define
\begin{equation}
\ti \Omega^{(2)}_{\Pinf,2r-2}=\begin{cases}
\sum_{q=0}^{r-1}  (2q+1)\tilde c_{r-1-q}\,
\omega^{(2)}_{\Pinf,2q} & \text{for
$r\in\bbN$}, \\
0 & \text{for $r=0$}. \end{cases}
\end{equation}
Then one infers
\begin{equation}
\int_{Q_0}^P
\ti \Omega^{(2)}_{\Pinf,2r-2}\underset{\zeta\to 0}{=}
\begin{cases}-\sum_{q=0}^{r-1} \tilde c_{r-1-q}
\zeta^{-2q-1}+\Oh(1)
 & \text{for $r\in\bbN$} \\
0 & \text{for $r=0$} \end{cases}\text{ as $P\to\Pinf$}.
\end{equation}
The corresponding  vector of $b$-periods
of $\ti \Omega^{(2)}_{\Pinf,2r-2} / (2\pi i)$ is then
denoted by
$\ul {\ti U}^{(2)}_{2r-2}$,
\begin{equation}
\ul {\ti U}_{2r-2}^{(2)}=
(\ti U_{2r-2,1}^{(2)},\dots,\ti U_{2r-2,n}^{(2)}), \quad
\ti U^{(2)}_{2r-2,j}=
\frac{1}{2\pi i} \int_{b_j} \ti \Omega^{(2)}_{\Pinf,2r-2},
\quad j=1,\dots,\N. \lb{4.91a}
\end{equation}
One computes from \eqref{a35a},
\begin{equation}
\ti U^{(2)}_{2r-2,j} =-2\sum^{r-1}_{q=0} \tilde c_{r-1-q}
\sum_{k=1}^\N c_j(k)c_{k-\N+q}(\ul E), \quad j=1,\dots,\N,
\lb{4.19m}
\end{equation}
with $c_k(\ul E)$ defined in \eqref{c}.
Moreover, let $\omega^{(2)}_{\hat P,0}$  be a normalized
differential of the second kind, holomorphic on
$\calK_\N\setminus \{\hatt P\}$, such that
\begin{equation}
\omega^{(2)}_{\hat P,0}=(\zeta^{-2}+
\Oh(1))d\zeta \text{ as $P\to \hatt P$}.
\end{equation}
In the special case $\hatt P=P_0$, the vector of
$b$-periods of
$(\tilde\alpha/\alpha)Q^{1/2}
\omega^{(2)}_{P_0,0} / (2\pi i)$ is
denoted by $\ul W_0^{(2)}$ and one obtains
(cf.\ \eqref{a35ak}),
\begin{equation}
\ul W_0^{(2)}=(W_{0,1}^{(2)},\dots,W_{0,n}^{(2)}), \quad
W^{(2)}_{0,j}=\frac{1}{2\pi i}
\frac{\tilde \alpha}{\alpha} Q^{1/2}
 \int_{b_j}\omega^{(2)}_{P_0,0}=
2\frac{\tilde \alpha}{\alpha} c_j(1),
\quad j=1,\dots,\N. \lb{4.91ab}
\end{equation}

\begin{theorem} \lb{theorem4.11}
Assume \eqref{3.14a}, \eqref{3.20kc},
$P\in\calK_\N\setminus\{\Pinf \}$
and $(x,t_\N), \, (x_0,t_{0,\N})\in\Omega$, where
$\Omega\subseteq\bbR^2$
is open and connected. Suppose that
$\calD_{\hat{\ul\mu}(x,t_r)}$,
or equivalently, $\calD_{\hat{\ul\nu}(x,t_r)}$ is nonspecial
for $(x,t_r)\in\Omega$. Then $\phi(P,x,t_r)$ admits the
representation
\begin{align}
\phi (P,x,t_r) = &\frac{\theta (\ul\Xi_{Q_0} - \ul A_{Q_0}
(\Pinf) + \ul\alpha_{Q_0} (\calD_{\hat{\ul\mu} (x,t_r)}))}
{\theta(\ul\Xi_{Q_0} -\ul A_{Q_0}
(\Pinf) +\ul\alpha_{Q_0} (\calD_{\hat{\ul\mu}(x,t_r)})+
\ul \Delta)} \times
\lb{4.30b}\\
& \times \frac{\theta (\ul\Xi_{Q_0} -
\ul A_{Q_0} (P) + \ul\alpha_{Q_0}
(\calD_{\hat{\ul\mu}(x,t_r)})+\ul \Delta)}{\theta (\ul\Xi_{Q_0}
-\ul A_{Q_0} (P) +
\ul\alpha_{Q_0}(\calD_{\hat{\ul\mu} (x,t_r)}))} \exp
\bigg(-\int_{Q_0}^P
\omega_{\Pinf,P_0}^{(3)}\bigg), \no
\end{align}
with the halfperiod $\ul \Delta$ defined in \eqref{3.24k}.
The components
$\psi_j(P,x,x_0,t_r,t_{0,r})$, $j=1,2$ of the Baker--Akhiezer
vector are given by
\begin{align}
&\psi_1(P,x,x_0,t_r,t_{0,r}) =
\frac{\theta (P,\hat{\ul\mu}(x,t_r))
\theta (\Pinf,\hat{\ul\mu}(x_0,t_{0,r}))}
{\theta (\Pinf,\hat{\ul\mu}(x,t_r))
\theta (P,\hat{\ul\mu}(x_0,t_{0,r}))}
  \exp \bigg(-i(x-x_0)\int_{Q_0}^P\omega_{\Pinf,0}^{(2)} \no \\
&\hspace*{1.5cm} +(t_r-t_{0,r})(\tilde \alpha/\alpha)Q^{1/2}
\int_{Q_0}^P\omega_{P_0,0}^{(2)}+
(t_r-t_{0,r})\int_{Q_0}^P\ti
\Omega_{\Pinf,2r-2}^{(2)}\bigg) \lb{4.31b}
\end{align}
and
\begin{align}
\psi_2(P,x,x_0,t_r,t_{0,r})
&= -\frac{\theta (P,\hat{\ul\mu}(x,t_r)),\ul \Delta)
\theta (\Pinf,\hat{\ul\mu}(x_0,t_{0,r}))}
{\theta (\Pinf,\hat{\ul\mu}(x,t_r)),\ul \Delta)
\theta (P,\hat{\ul\mu}(x_0,t_{0,r}))} \exp \bigg(-\int_{Q_0}^P
\omega_{\Pinf,P_0}^{(3)} \no \\
& \quad -i(x-x_0)\int_{Q_0}^P
\omega_{\Pinf,0}^{(2)} +(t_r-t_{0,r})
(\tilde \alpha/\alpha)Q^{1/2} \int_{Q_0}^P
\omega_{P_0,0}^{(2)} \no \\
& \quad +(t_r-t_{0,r})\int_{Q_0}^P
\ti \Omega_{\Pinf,2r-2}^{(2)}\bigg).
\lb{4.32}
\end{align}
The Abel map linearizes the auxiliary divisors in the
sense that
\begin{subequations} \lb{4.33c}
\begin{align}
\ul\alpha_{Q_0} (\calD_{\hat{\ul\mu}(x,t_r)}) &=
\ul\alpha_{Q_0} (\calD_{\hat{\ul\mu}(x_0,t_{0,r})})
-2 i\ul c(\N)(x-x_0) + \ul d_r (t_r-t_{0,r}),
\lb{4.33ca} \\
\ul\alpha_{Q_0} (\calD_{\hat{\ul\nu}(x,t_r)}) &=
\ul\alpha_{Q_0} (\calD_{\hat{\ul\nu}(x_0,t_{0,r})})
-2 i\ul c(\N)(x-x_0) + \ul d_r (t_r-t_{0,r}),
\lb{4.33cb}
\end{align}
\end{subequations}
where $\ul c(\N)$ is defined in \eqref{A.7} and
$\ul d_r$ is given by
\begin{equation}
\ul d_r=\begin{cases}-2 (\tilde\alpha / \alpha)\ul c(1)
+2 \sum_{q=0}^{r-1}\tilde c_{r-1-q}
\sum_{k=1}^\N \ul c(k) \, c_{k-\N+q}(\ul E) & \text{for
$r\in\bbN$}, \\
-2 (\tilde \alpha / \alpha) \ul c(1) & \text{for $r=0$},
\end{cases}  \lb{4.33s}
\end{equation}
with $c_k(\ul E)$ defined in \eqref{c} and the convention
$c_{-k}(\ul E)=0$
for $k\in\bbN$.

The solution
$u(x,t_r)$ of equations \eqref{4.1} and
\eqref{4.2} reads
\begin{align}
u(x,t_r)&=u(x_0,t_{r}) \lb{4.34b}\\
&\quad +2i\ln\left(
\f{\theta (\Pinf,\hat{\ul\mu}(x,t_r),\ul \Delta) \theta
(\Pinf,\hat{\ul\mu}(x_0,t_{0,r}))}
{\theta (\Pinf,\hat{\ul\mu}(x_0,t_{0,r}),\ul \Delta) \theta
(\Pinf,\hat{\ul\mu}(x,t_r)) }
\exp \big(- i e_0(x-x_0)\big) \right),\no
\end{align}
where we had to resort to the abbreviations \eqref{3.24d}.
\end{theorem}

\begin{proof}
The proofs of \eqref{4.30b} and \eqref{4.34b} carry over
without
changes from the stationary situation described in Theorem
\ref{theorem3.3} since they are based on the Riccati-type
equation \eqref{4.13a} in both cases, with essentially
the same
expression for $\phi$. Moreover, that the constructed
$u(x,t_r)$
indeed satisfies $\sG_r(u)=0$ and the $\N$th stationary
sGmKdV equation readily follows from
\begin{equation}
(U_{t_r}- \ti V_{r,x} +[U, \ti V_r])\Psi=\Psi_{xt_r}
-\Psi_{t_rx}=0
\end{equation}
and
\begin{equation}
(V_{\N,x}-[U,V_\N])\Psi=0
\end{equation}
applying \eqref{4.14a}--\eqref{4.14b} repeatedly.
Hence we turn to the
Baker--Akhiezer function $\Psi$ whose first component
$\psi_1$ is given by \eqref{4.24a},
\begin{align}
&\psi_1(P,x,x_0,t_r,t_{0,r}) \lb{4.34c}\\
&=\bigg(\f{F_\N(z,x,t_r)}{F_\N(z,x_0,t_{0,r})}\bigg)^{1/2}
\exp\bigg(iy(P)\int_{x_0}^x \f{dx'}{F_\N(z,x',t_r)}
 -\f{y(P)}{z}\int_{t_{0,r}}^{t_r}ds\, \f{\ti
F_r(z,x_0,s)}{F_\N(z,x_0,s)}\bigg). \no
\end{align}
The time-dependent term in the exponential has two potential
singularities,
one at $\Pinf$, and
the other at $P_0$. We first study this term  as
$P\to\Pinf$ (using the local coordinate
$\zeta=\sigma /z^{1/2}\to 0$, $\sigma =\pm 1$,
cf.\ \eqref{3.19dc}).
First assume that $\ti F_r$
is homogeneous, that is,
$\ti F_r=\hatt F_r$. Then, using the high-energy expansion
\eqref{high} at
$x=x_0$, equation
\eqref{2.19} at $j=r$,
 as well as \eqref{2.17aab} (with $\N$ replaced by $r$,
$\alpha$ by $\tilde \alpha$, and $\beta$ by $\tilde \beta$),
\begin{align}
\f{y(P)\hatt F_r(z,x_0,t_r)}{z F_\N(z,x_0,t_r)}&=
\zeta^2\bigg(\sum_{q=0}^r \hat f_{r-q}(x_0,t_r)z^q\bigg)
\f{y(P)}{F_\N(z,x_0,t_r)} \\
&=\zeta^{-2r+1}+(\tilde f_r(x_0,t_r)-\hat f_r(x_0,t_r))\zeta
+\Oh(\zeta^3) \text{ as $P\to\Pinf$}. \no
\end{align}
Here $\hat f_r$ is computed from the
KdV recursion \eqref{2.15a}, whereas $\tilde f_r=\tilde
\alpha e^{-iu}$.
Similarly,
\begin{equation}
\frac{y(P)\hatt F_q(z,x_0,t_r)}
{zF_\N(z,x_0,t_r)}\underset{\zeta\to 0}{=}
\zeta^{-2q+1} + \Oh(\zeta^3) \text{ as } P\to\Pinf, \quad
q=0,\dots,r-1.
\end{equation}
Hence one concludes
\begin{align}
\f{y(P)\ti F_r(z,x_0,t_r)}{z F_\N(z,x_0,t_r)}&=
\sum_{q=0}^r \tilde c_{r-q}\f{y(P) \hatt F_q(z,x_0,t_r)}
{z F_\N(z,x_0,t_r)} \no \\
& \underset{\zeta\to 0}{=} \begin{cases}
\sum_{q=1}^r \tilde c_{r-q}\zeta^{-2q+1}+\Oh(\zeta)
& \text{for $r\in\bbN$}, \\
\zeta + \Oh(\zeta^3) & \text{for $r=0$} \end{cases}
\end{align}
and thus
\begin{align} \lb{4.86k}
& \f{y(P)}{z} \int_{t_{0,r}}^{t_r} ds\,  \f{\ti F_r(z,x_0,s)}
{F_\N(z,x_0,s)} \no \\
&\underset{\zeta\to 0}{=}
\begin{cases}(t_r-t_{0,r})\big(\sum_{q=0}^{r-1}
\tilde c_{r-1-q}\zeta^{-2q-1}
\big)+\Oh(\zeta) &
\text{for $r\in\bbN$}, \\
(t_r-t_{0,r})\zeta+\Oh(\zeta^3) & \text{for $r=0$}, \end{cases}
\text{ as } P \to \Pinf.
\end{align}
Secondly, we study the behavior near $P_0$ (using the local
coordinate $\zeta=\sigma z^{1/2} \to 0$, $\sigma = \pm 1$,
cf.\ \eqref{3.19dc}). One finds
\begin{equation}
\f{y(P)\ti F_r(z,x_0,t_r)}{z F_\N(z,x_0,t_r)}
\underset{\zeta\to 0}{=} \f{\tilde \alpha}{\alpha}Q^{1/2}
\zeta^{-1}+\Oh(\zeta) \text{ as } P \to P_0
\end{equation}
and hence
\begin{equation} \lb{4.88k}
\f{y(P)}{z}\int_{t_{0,r}}^{t_r}ds\, \f{\ti F_r(z,x_0,s)}
{F_\N(z,x_0,s)}\underset{\zeta\to 0}{=} \f{\tilde \alpha}{\alpha}
Q^{1/2}(t_r-t_{0,r})\zeta^{-1}+\Oh(\zeta) \text{ as } P \to P_0.
\end{equation}
Here the sign of $Q^{1/2}$ is determined by the
compatibility of
the charts near $\Pinf$ and $P_0$.
Together with  our choice of signs in \eqref{b27ac} one
obtains
\begin{equation}
Q^{1/2}=(-1)^\N \prod_{m=1}^{2\N}|E_m^{1/2}|.
\end{equation}
Equations \eqref{4.34c}, \eqref{4.86k}, and \eqref{4.88k}
then yield
\eqref{4.31b} as in Theorem \ref{theorem3.3}.

Next we indicate the proof of \eqref{4.33ca} using
Lemma \ref{lemma4.6} and following
the corresponding argument in the proof of Theorem
\ref{theorem3.3}.
We temporarily assume,
\begin{equation}
\mu_j(x,t_r)\neq \mu_{j'}(x,t_r) \text{ for } j\neq j'
\text{ and } (x,t_r)\in\ti \Omega\subseteq\Omega, \lb{4.88r}
\end{equation}
where $\ti \Omega$ is open and connected. Equations
\eqref{4.26a},
\eqref{symm},
Lagrange's interpolation theorem, Theorem \ref{theoremB1}, and
\eqref{a35a} then yield for $r\in\bbN$,
\begin{align}
&\f{\partial}{\partial t_r}
\ul\alpha_{Q_0}(\calD_{\hat{\ul\mu}(x,t_r)})=\sum_{j=1}^\N
\mu_{j,t_r}(x,t_r)\sum_{k=1}^\N\ul
c(k)\f{\mu_j(x,t_r)^{k-1}}{y(\hat\mu_j(x,t_r))}
\no \\
&=2\sum_{j,k=1}^\N \ul c(k) \f{\mu_j(x,t_r)^{k-1}}
{\prod_{\substack{\ell=1\\ \ell\neq j}}(\mu_j(x,t_r)-
\mu_\ell(x,t_r))}\,
\f{\ti F_r(\mu_j(x,t_r),x,t_r)}{\mu_j(x,t_r)}\no \\
&=2\sum_{j,k=1}^\N \ul c(k) \f{\mu_j(x,t_r)^{k-1}}
{\prod_{\substack{\ell=1\\ \ell\neq j}}(\mu_j(x,t_r)-
\mu_\ell(x,t_r))}
\bigg(\sum_{q=0}^{r-1}
\sum_{p=(q-\N)\maxi 0}^q \tilde c_{r-1-q}\, c_p(\ul E)
\Phi_{q-p}^{(j)}(\ul\mu) \no \\
&\hspace*{9.5cm} -
\f{\tilde\alpha}{\alpha}\Phi_{\N-1}^{(j)}(\ul\mu)\bigg)\no \\
&=2\sum_{k=1}^\N \ul c(k)\bigg(\sum_{q=0}^{r-1}
\sum_{p=(q-\N)\maxi 0}^q
\tilde c_{r-1-q}\, c_p(\ul E)\sum_{j=1}^\N
\f{\mu_j(x,t_r)^{k-1}}{\prod_{\substack{\ell=1\\ \ell\neq
j}}(\mu_j(x,t_r)-\mu_\ell(x,t_r))}
\Phi_{q-p}^{(j)}(\ul\mu) \no \\
&\hspace*{5cm} -\f{\tilde\alpha}{\alpha}\sum_{j=1}^\N
\f{\mu_j(x,t_r)^{k-1}}{\prod_{\substack{\ell=1\\ \ell\neq
j}}(\mu_j(x,t_r)-\mu_\ell(x,t_r))}
\Phi_{\N-1}^{(j)}(\ul\mu)\bigg)\no \\
&=2\sum_{k=1}^\N \ul c(k)\bigg(\sum_{q=0}^{r-1}
\sum_{p=(q-\N)\maxi 0}^q \tilde c_{r-1-q}\,
c_p(\ul E)\delta_{k,\N-(q-p)}-
\f{\tilde\alpha}{\alpha}\delta_{k,1}\bigg) \no \\
&=\ul d_r =-(\ul {\ti U}^{(2)}_{2r-2} +{\ul W}^{(2)}_0) \lb{4.100}
\end{align}
and hence \eqref{4.33ca} subject to \eqref{4.88r}. The extension
from $(x,t_r)\in\ti \Omega$ to $(x,t_r)\in\Omega$ then follows
by continuity of $\ul \alpha_{Q_0}$ as in Theorem
\ref{theorem3.3}. \eqref{4.33cb} then
follows from
\eqref{4.33ca} and the linear equivalence of
$\calD_{\Pinf \hat {\ul \mu}(x,t_r)}$ and
$\calD_{P_0 \hat {\ul \nu}(x,t_r)}$, that is,
\begin{equation}
\ul\alpha_{Q_0}(\calD_{\hat{\ul\nu}(x,t_r)})=
\ul\alpha_{Q_0}(\calD_{\hat{\ul\mu}(x,t_r)})+ \ul \Delta.
\lb{4.34p}
\end{equation}
\end{proof}

\begin{remark} \lb{remark4.12}
Equations \eqref{4.31b}--\eqref{4.34b} have been derived
for the sG equation (and its close allies), that is, in
the case $r=1$, by a variety of
authors, we refer, for instance, to \cite{BBEIM94},
Sects.\ 4.2, 4.3, \cite{Ch82}, \cite{Da82},
\cite{EF85}, \cite{KK76}, \cite{Ma76}, and \cite{TCL87}.
Moreover, considerable efforts have been devoted to
the construction of real-valued solutions by imposing certain
symmetry conditions on the (nonzero) branch points
$(E_m,0)$, $m=1,\dots,2\N$, see \cite{BBEIM94}, Sect.\ 4.3.3,
\cite{Bo91a},
\cite{Bo91b}, \cite{Da82}, \cite{Du82b}, \cite{DN82a},
\cite{DN82b}, \cite{Er89}, \cite{EF85}, \cite{EFM86a},
\cite{EKT93}, \cite{FM82}, \cite{FM83}, \cite{No85},
\cite{PS89}, \cite{Sm91}, \cite{Ta90a}, \cite{TTCL84},
and the references therein.
\end{remark}

\begin{remark} \lb{remark4.13}
The linearization property \eqref{4.33ca} (and \eqref{3.23aa})
can also be
obtained via an alternative procedure which we briefly sketch.
Following a standard argument (see, e.g.,
\cite{NMPZ84}, p.\ 139--144) one introduces the meromorphic
differential
\begin{equation}
\Omega_1(x,x_0,t_r,t_{0,r})= \frac{\partial}{\partial z}
\ln (\psi_1(\dott,x,x_0,t_r,t_{0,r})) dz
\end{equation}
and hence infers from the representation \eqref{4.31b}
\begin{align}
\Omega_1(x,x_0,t_r,t_{0,r})=&-i(x-x_0)\omega^{(2)}_{\Pinf,0}
+ (t_r-t_{0,r})\ti \Omega^{(2)}_{\Pinf,2r-2} +(t_r-t_{0,r})
\frac{\tilde \alpha}{\alpha}Q^{1/2}\omega^{(2)}_{P_0,0} \no
\\ & \quad - \sum_{j=1}^\N \omega^{(3)}_{\hat { \mu}_j(x_0,t_{0,r}),
\hat { \mu}_j(x,t_r)} + \omega. \lb{4.88b}
\end{align}
Here $\omega$ denotes a holomorphic differential on $\calK_\N$,
that is,
\begin{equation}
\omega=\sum_{j=1}^\N c_j \omega_j \lb{3.49c}
\end{equation}
for some $c_j\in\bbC$, $j=1,\dots,\N$. Since
$\psi_1(\dott,x,x_0,t_r,t_{0,r})$
is single-valued on $\calK_\N$, all $a$ and $b$-periods of
$\Omega_1$ are integer multiples of $2\pi i$ and hence
\begin{equation}
2\pi i m_k= \int_{a_k} \Omega_1(x,x_0,t_r,t_{0,r})=\int_{a_k}
\omega =c_k,
\quad j=1,\dots,\N \lb{3.49d}
\end{equation}
for some $m_k\in\bbZ$ identifies $c_k$ as integer multiples of
$2\pi i$. Similarly, for some $n_k\in\bbZ$,
\begin{align}
2\pi in_k &=\int_{b_k} \Omega_1(x,x_0,t_r,t_{0,r})  \no \\
 &=-i(x-x_0)\int_{b_k}
\omega^{(2)}_{\Pinf,0} + (t_r-t_{0,r})
\frac{\tilde \alpha}{\alpha}
Q^{1/2}\int_{b_k}\omega^{(2)}_{P_0,0} \no \\
& \quad +(t_r-t_{0,r}) \int_{b_k} \ti \Omega^{(2)}_{\Pinf,2r-2}
 -\sum_{j=1}^\N \int_{b_k}
\omega^{(3)}_{\hat { \mu}_j(x_0,t_{0,r}), \hat { \mu}_j(x,t_r)}
+ 2\pi i \sum_{j=1}^\N m_j \int_{b_k} \omega_j  \no \\
&=2\pi U^{(2)}_{0,k} (x-x_0) +2\pi (\ti U_{2r-2,k}^{(2)}
+  W_{0,k}^{(2)})(t_r-t_{0,r}) \no \\
& \quad -2\pi i \sum_{j=1}^\N
\big(\ul {A}_{\hat \mu_j(x,t_r)}(\hat \mu_j(x_0,t_{0,r})\big)_k
+2\pi i \sum_{j=1}^\N m_j \tau_{j,k}, \quad k=1,\dots,\N,
\lb{3.49e}
\end{align}
using \eqref{3.49ea}, \eqref{4.91a}, \eqref{4.91ab}, and
\eqref{a37}.
By symmetry of $\tau$ (cf.\ \eqref{a18a}),
\eqref{3.49e} is equivalent to
\begin{equation}
{\ul \alpha}_{Q_0}(\calD_{\hat {\ul \mu}(x,t_r)})
= {\ul \alpha}_{Q_0}(\calD_{\hat {\ul \mu}(x_0,t_{0,r})})
+i{\ul U}^{(2)}_0 (x-x_0) -(\ul {\ti U}_{2r-2}^{(2)} +
\ul W_0^{(2)})(t_r-t_{0,r}). \lb{3.49f}
\end{equation}
\end{remark}

Finally, we remark on an interesting property of the
time-dependent sGmKdV hierarchy in connection with its
algebro-geometric solutions. In fact, equations \eqref{4.1}
and \eqref{4.2} can be rewritten so that $u(x,t_r)$ satisfies
a differential
equation with a pure first-order time derivative.

\begin{remark} \lb{remark4.8}
The solution $u$ of equations \eqref{4.1} and \eqref{4.2}
satisfies
\begin{equation}
u_{t_r}=2i \alpha^{-1} (\tilde \alpha g_{\N-1}-
\alpha \tilde g_{r-1}),
\quad u(x,t_{0,r})=u^{(0)}(x).
\lb{4.19a}
\end{equation}
Indeed,  \eqref{2.24} yields
\begin{equation}
f_\N=\alpha e^{-iu}, \quad \tilde f_r=\tilde \alpha e^{-iu},
\lb{4.19b}
\end{equation}
which inserted into the constant term (i.e., the coefficient
of $z^0$) in \eqref{4.18a} results in
\begin{equation}
-i\alpha u_{t_r} e^{-iu}=2(g_{\N-1}
\tilde f_r-\tilde g_{r-1}f_\N). \lb{4.19c}
\end{equation}
\end{remark}

Remark \ref{remark4.8} can be illustrated as follows. \\
(i) Consider $\N=1$. \\
(ia) $r=0$. Then \eqref{4.19a} becomes
\begin{equation}
u_{t_0}+i\f{\tilde\alpha}{\alpha}u_x=0,
\end{equation}
with solution
\begin{equation}
u(x,t_0)=u^{(0)}(x-i\f{\tilde\alpha}{\alpha} t_0).
\end{equation}
(ib) $r=1$. Then \eqref{4.19a} reads
\begin{equation}
u_{t_1}+i(\f{\tilde\alpha}{\alpha}-1)u_x=0,
\end{equation}
with solution
\begin{equation}
u(x,t_1)=u^{(0)}(x-i(\f{\tilde\alpha}{\alpha}-1)t_1).
\end{equation}
(ii) Consider $\N=2$. \\
(iia) $r=0$. Then \eqref{4.19a} becomes
\begin{equation}
u_{t_0}=i\f{\tilde\alpha}{8\alpha}(u^3_x+2u_{xxx})
-i\f{\tilde\alpha}{\alpha}c_1 u_x,\quad
u(x,t_{0,0})=u^{(0)}(x).
\end{equation}
(iib) $r=1$. Then \eqref{4.19a} reads
\begin{equation}
u_{t_1}=i\f{\tilde\alpha}{8\alpha}(u^3_x+2u_{xxx})
+i(1-c_1 \f{\tilde\alpha}{8\alpha}) u_x,\quad
u(x,t_{0,1})=u^{(0)}(x).
\end{equation}
(iic) $r=2$. Then \eqref{4.19a} becomes
\begin{equation}
u_{t_2}=-\f{i}{8}(\f{\tilde\alpha}{\alpha}-1)(u_x^3+2u_{xxx})+
i(\tilde c_1-\f{\tilde\alpha}{\alpha}c_1) u_x,\quad
u(x,t_{0,2})=u^{(0)}(x).
\end{equation}

\appendix
\section{Hyperelliptic curves and their theta functions}\lb{A}
\renewcommand{\theequation}{A.\arabic{equation}}
\renewcommand{\thetheorem}{A.\arabic{theorem}}
\setcounter{theorem}{0}
\setcounter{equation}{0}

We give a brief summary of some of the fundamental properties
and notations needed from the
theory of hyperelliptic curves.  More details can be found in
some of the standard textbooks
\cite{FK92} and \cite{Mu84}, as well as monographs
dedicated to integrable systems such as \cite{BBEIM94}, Ch.\ 2,
\cite{GH98}, App. A--C, and \cite{NMPZ84}.

Fix $\N\in\bbN$. The hyperelliptic curve $\calK_\N$
of genus $\N$ is defined by
\begin{multline}
\calK_\N: \, \calF_\N(z,y)=y^2-R_{2\N+1}(z)=0, \quad
R_{2\N+1}(z)=\prod_{m=0}^{2\N}(z-E_m), \\
\quad E_0=0, \quad E_1,\dots,E_{2\N}\in\bbC, \quad
E_m \neq E_{n} \text{ for } m \neq n, \, m,n=0,\dots,2\N.
\label{b1}
\end{multline}
The curve \eqref{b1} is compactified by adding one
point $\Pinf$ at infinity.
One introduces an appropriate set of
$\N$ nonintersecting cuts $\calC_j$ joining
$E_{m(j)}$ and $E_{m^\prime(j)}$ and $\calC_\infty$
joining $E_{2\N}$ and
$\infty$.
Denote
\begin{equation}
\calC=\bigcup_{j\in \{1,\dots,\N \}\cup\{\infty\}}\calC_j,
\quad
\calC_j\cap\calC_k=\emptyset,
\quad j\neq k.\label{b2}
\end{equation}
Define the cut plane
\begin{equation}
\Pi=\bbC\setminus\calC, \label{b3}
\end{equation}
and introduce the holomorphic function
\begin{equation}
R_{2\N+1}(\dott)^{1/2}\colon \Pi\to\bbC, \quad
z\mapsto \left(\prod_{m=0}^{2\N}(z-E_m) \right)^{1/2}\label{b4}
\end{equation}
on $\Pi$ with an appropriate choice of the square root
branch in \eqref{b4}.
Define
\begin{equation}
\calM_{\N}=\{(z,\sigma R_{2\N+1}(z)^{1/2}) \mid
z\in\bbC,\; \sigma\in\{\pm 1\}
\}\cup\{\Pinf\} \label{b5}
\end{equation}
by extending $R_{2\N+1}(\dott)^{1/2}$ to $\calC$. The
hyperelliptic curve
$\calK_\N$ is then the set
$\calM_{\N}$ with its natural complex structure obtained
upon gluing the
two sheets of $\calM_{\N}$
crosswise along the cuts. Finite points
$P$ on $\calK_\N$ are denoted by
$P=(z,y)$, where $y(P)$ denotes the meromorphic function
on $\calK_\N$
satisfying $\calF_\N(z,y)=y^2-R_{2\N+1}(z)=0$;
$\calK_\N$ has  genus $\N$.

One verifies that $dz/y$ is a holomorphic differential
on $\calK_\N$ with zeros of order $2(\N-1)$ at  $\Pinf$
and hence
\begin{equation}
\eta_j=\frac{z^{j-1}dz}{y}, \quad j=1,\dots,\N
\lb{b24}
\end{equation}
form a basis for the space of holomorphic differentials
on $\calK_\N$.  Introducing the
invertible matrix $C$ in $\bbC^\N$,
\begin{align}
\begin{split}
C & =(C_{j,k})_{j,k=1,\dots,\N}, \quad C_{j,k}
= \int_{a_k} \eta_j, \\
\underline{c} (k) & = (c_1(k), \dots,
c_n(k)), \quad c_j (k) =
C_{j,k}^{-1},
\lb{A.7}
\end{split}
\end{align}
the normalized differentials $\omega_j$ for $j=1,\dots,\N$,
\begin{equation}
\omega_j = \sum_{\ell=1}^\N c_j (\ell) \eta_\ell,
\quad \int_{a_k} \omega_j =
\delta_{j,k}, \quad j,k=1,\dots,\N
\lb{b26}
\end{equation}
form a canonical basis for the space of
holomorphic differentials on $\calK_\N$.
Here $\{a_j,b_j\}_{j=1,\dots,\N}$ is a homology basis for
$\calK_\N$.
Near $\Pinf$ one infers
\begin{align}
\ul\omega & = (\omega_1,\dots,\omega_\N)=
-2 \bigg( \sum_{j=1}^\N \f{\ul c (j)
\zeta^{2(\N-j)}}{\big(\prod_{m=0}^{2\N}
(1-E_m \zeta^2) \big)^{1/2}} \bigg) d\zeta \no \\
& \underset{\zeta \to 0}{=}
-2\bigg( \sum_{q=0}^{\infty}\sum_{k=1}^\N
\ul c(k)c_{k-\N+q}(\ul E)\zeta^{2q}\bigg) d\zeta \lb{b27k} \\
& \underset{\zeta \to 0}{=} -2 \bigg( \ul c (\N) +
\big( \frac12 \ul c (\N)
\sum_{m=0}^{2\N} E_m +\ul c
(\N-1) \big) \zeta^2 + \Oh(\zeta^4) \bigg)d\zeta, \lb{b27} \\
& \hspace*{5.5cm}  \zeta=\sigma /z^{1/2},
\quad \sigma=\pm 1, \no
\end{align}
with $c_k(\ul E)$ defined in \eqref{c} and
\begin{equation}
y(P) \underset{\zeta \to 0}{=} \bigg(1-\frac12
\bigg( \sum_{m=0}^{2\N} E_m \bigg)\zeta^2 +
\Oh(\zeta^4)\bigg)\zeta^{-2\N-1} \text{ as }
P\to\Pinf. \lb{b27a}
\end{equation}
Similarly, near $P_0=(0,0)$ one computes
\begin{align}
&\ul\omega \underset{\zeta\to 0}{=} 2 \bigg( \frac{\ul c(1)}
{Q^{1/2}} + \Oh(\zeta^2) \bigg)d\zeta, \quad Q^{1/2}=
(-1)^\N \prod_{m=1}^{2\N}|E_m^{1/2}|, \lb{b27ab} \\
& \hspace*{5.5cm} \zeta=\sigma z^{1/2}, \quad \sigma =\pm 1,
\no
\end{align}
using
\begin{equation}
y(P)\underset{\zeta\to 0}{=} Q^{1/2}\zeta + \Oh(\zeta^3)
\text{ as } P\to P_0. \lb{b27ac}
\end{equation}
Associated with the homology basis
$\{a_j, b_j\}_{j=1,\dots,\N}$ we
also recall the canonical dissection of $\calK_\N$
along its cycles yielding
the simply connected interior $\hatt \calK_\N$ of the
fundamental polygon $\partial {\hatt \calK}_\N$ given by
\begin{equation}
\partial  {\hatt \calK}_\N =a_1 b_1 a_1^{-1} b_1^{-1}
a_2 b_2 a_2^{-1} b_2^{-1} \cdots
a_\N^{-1} b_\N^{-1}.
\lb{a25}
\end{equation}
Let $\calM (\calK_\N)$ and $\calM^1 (\calK_\N)$ denote the
set of meromorphic
functions (0-forms) and meromorphic
differentials (1-forms)
on $\calK_\N$. The residue of a meromorphic differential
$\nu\in \calM^1 (\calK_\N)$ at a
point $Q \in \calK_\N$ is defined by
\begin{equation}
\text{res}_{Q}(\nu)
=\frac{1}{2\pi i} \int_{\gamma_{Q}} \nu,
\lb{a33}
\end{equation}
where $\gamma_{Q}$ is a counterclockwise oriented
smooth simple closed
contour encircling $Q$ but no other pole of
$\nu$.  Holomorphic
differentials are also called Abelian differentials
of the first kind (dfk). Abelian differentials of the
second kind
(dsk) $\omega^{(2)} \in \calM^1 (\calK_\N)$ are characterized
by the property
that all their residues vanish.  They are
normalized, for
instance, by demanding that all their $a$-periods
vanish, that is,
\begin{equation}
\int_{a_j} \omega^{(2)} =0, \quad  j=1,\dots,\N.
\lb{a34}
\end{equation}
If $\omega_{P_1, n}^{(2)}$ is a dsk on $\calK_\N$ whose
only pole is $P_1 \in \hatt \calK_\N$ with principal part
$\zeta^{-n-2}\,d\zeta$, $n\in\bbN_0$ near
$P_1$ and $\omega_j =
 (\sum_{m=0}^\infty d_{j,m} (P_1) \zeta^m)\, d\zeta$
near $P_1$, then
\begin{equation}
\int_{b_j} \omega_{P_1, n}^{(2)} =
 \frac{2\pi i}{n+1} d_{j,n} (P_1).
\lb{a35}
\end{equation}
In particular,
\begin{align}
& \f{1}{2\pi i}\int_{b_j}\omega^{(2)}_{\Pinf,0}=-2 c_j(\N),
\no \\
& \f{2q+1}{2\pi i}\int_{b_j}\omega^{(2)}_{\Pinf,2q}
=-2\sum_{k=1}^\N c_j(k)c_{k-\N+q}(\ul E),
\quad j=1,\dots,\N, \,\, q\in\bbN_0, \lb{a35a} \\
& \f{1}{2\pi i}\int_{b_j}\omega^{(2)}_{P_0,0}
=2 \frac{c_j(1)}{Q^{1/2}}, \quad j=1,\dots,\N,
\lb{a35ak}
\end{align}
using \eqref{b27k} and \eqref{b27ab}.

Any meromorphic differential $\omega^{(3)}$ on
$\calK_\N$ not of the first or
second kind is said to be of the third
kind (dtk).
A dtk $\omega^{(3)} \in \calM^1 (\calK_\N)$
is usually normalized by the vanishing of its
$a$-periods, that is,
\begin{equation}
\int_{a_j} \omega^{(3)} =0, \quad  j=1,\dots, \N.
\lb{a36}
\end{equation}
A normal dtk $\omega_{P_1, P_2}^{(3)}$ associated
with two points $P_1$,
$P_2 \in \hatt \calK_\N$, $P_1 \neq P_2$ by definition
has simple poles at
$P_j$ with residues $(-1)^{j+1}$, $j=1,2$ and
vanishing $a$-periods.  If $\omega_{P,Q}^{(3)}$ is a
normal dtk associated
with $P$, $Q\in\hatt \calK_\N$, holomorphic on
$\calK_\N \setminus \{ P,Q\}$, then
\begin{equation}
\int_{b_j} \omega_{P,Q}^{(3)} =2\pi i \int_{Q}^P \omega_j,
\quad  j=1,\dots,\N,
\lb{a37}
\end{equation}
where the path from $Q$ to $P$ lies in
$\hatt \calK_\N$ (i.e.,
does not touch any of the cycles $a_j$, $b_j$).

We shall always assume (without loss of generality)
that all poles of
dsk's and dtk's on $\calK_\N$ lie on $\hatt \calK_\N$ (i.e.,
not on $\partial \hatt \calK_n$).

Define the matrix $\tau=(\tau_{j,\ell})_{j,\ell=1,\dots,\N}$ by
\begin{equation}
\tau_{j,\ell}=\int_{b_j}\omega_\ell, \quad j,\ell=1,
\dots,\N. \label{b8}
\end{equation}
Then
\begin{equation}
\Im(\tau)>0, \quad \text{and} \quad \tau_{j,\ell}=\tau_{\ell,j},
\quad j,\ell =1,\dots,\N.  \lb{a18a}
\end{equation}
Associated
with $\tau$ one introduces the period lattice
\begin{equation}
L_\N = \{ \ul z \in\bbC^\N \mid \ul z = \ul m +\tau \ul n,
\; \ul m, \ul n \in\bbZ^\N\}
\lb{a28}
\end{equation}
and the Riemann theta function associated with $\calK_\N$ and
the given homology basis $\{a_j,b_j\}_{j=1,\dots,\N}$,
\begin{equation}
\theta(\ul z)=\sum_{\ul n\in\bbZ^\N}\exp\big(2\pi
i(\ul n,\ul z)+\pi
i(\ul n,\tau \ul n)\big),
\quad \ul z\in\bbC^\N, \label{b9}
\end{equation}
where $(\ul u, \ul v)=\sum_{j=1}^\N \overline{u}_j v_j$
denotes the
scalar product
in $\bbC^\N$. It has the fundamental properties
\begin{align}
& \theta(z_1, \ldots, z_{j-1}, -z_j, z_{j+1},
\ldots, z_\N) =\theta
(\ul z), \lb{a27}\\
& \theta (\ul z +\ul m +\tau \ul n)
=\exp \big(-2 \pi i (\ul n,\ul z) -\pi i (\ul n, \tau
\ul n) \big) \theta (\ul z), \quad \ul m, \ul n \in\bbZ^\N.
\lb{aa51}
\end{align}

Next, fix a base point $Q_0\in\calK_\N\setminus\{P_0,\Pinf\}$,
denote by
$J(\calK_\N) = \bbC^\N/L_\N$ the Jacobi variety of $\calK_\N$,
and define the
Abel map $\underline{A}_{Q_0}$ by
\begin{equation}
\underline{A}_{Q_0} \colon \calK_\N \to J(\calK_\N), \quad
\underline{A}_{Q_0}(P)=
\big(\int_{Q_0}^P \omega_1,\dots,\int_{Q_0}^P \omega_\N \big)
\pmod{L_\N}, \, P\in\calK_\N. \label{b10}
\end{equation}
Similarly, we introduce
\begin{equation}
\ul \alpha_{Q_0}  \colon
\Div(\calK_\N) \to J(\calK_\N),\quad
\calD \mapsto \ul \alpha_{Q_0} (\calD)
=\sum_{P \in \calK_\N} \calD (P) \ul A_{Q_0} (P),
\label{aa47}
\end{equation}
where $\Div(\calK_\N)$ denotes the set of
divisors on $\calK_\N$. Here $\calD \colon \calK_\N \to \bbZ$
is called a divisor on $\calK_\N$ if $\calD(P)\neq0$ for only
finitely many $P\in\calK_\N$.

In connection with divisors on $\calK_\N$ we shall employ the
following
(additive) notation,
\begin{multline} \lb{A.17}
\calD_{Q_0\ul Q}=\calD_{Q_0}+\calD_{\ul Q}, \quad \calD_{\ul
Q}=\calD_{Q_1}+\cdots +\calD_{Q_n}, \\
  {\ul Q}=(Q_1, \dots ,Q_n) \in \sigma^n \calK_\N,
\quad Q_0\in\calK_\N,
\end{multline}
where for any $Q\in\calK_\N$,
\begin{equation} \lb{A.18}
\calD_Q \colon  \calK_\N \to\bbN_0, \quad
P \mapsto  \calD_Q (P)=
\begin{cases} 1 & \text{for $P=Q$},\\
0 & \text{for $P\in \calK_\N\setminus \{Q\}$}, \end{cases}
\end{equation}
and $\sigma^n \calK_\N$ denotes the $n$th symmetric product of
$\calK_\N$. In particular, $\sigma^m \calK_\N$ can be
identified with
the set of nonegative
divisors $0 \leq \calD \in \Div(\calK_\N)$ of degree $m$.

For $f\in \calM (\calK_\N) \setminus \{0\}$,
$\omega \in \calM^1 (\calK_\N) \setminus \{0\}$ the
divisors of $f$ and $\omega$ are denoted
by $(f)$ and
$(\omega)$, respectively.  Two
divisors $\calD$, $\calE\in \Div(\calK_\N)$ are
called equivalent, denoted by
$\calD \sim \calE$, if and only if $\calD -\calE
=(f)$ for some
$f\in\calM (\calK_\N) \setminus \{0\}$.  The divisor class
$[\calD]$ of $\calD$ is
then given by $[\calD]
=\{\calE \in \Div(\calK_\N)\mid\calE \sim \calD\}$.  We
recall that
\begin{equation}
\deg ((f))=0,\, \deg ((\omega)) =2(\N-1),\,
f\in\calM (\calK_\N) \setminus
\{0\},\,  \omega\in \calM^1 (\calK_\N) \setminus \{0\},
\lb{a38}
\end{equation}
where the degree $\deg (\calD)$ of $\calD$ is given
by $\deg (\calD)
=\sum_{P\in \calK_\N} \calD (P)$.  It is custom to call
$(f)$ (respectively,
$(\omega)$) a principal (respectively, canonical)
divisor.

Introducing the complex linear spaces
\begin{align}
\calL (\calD) & =\{f\in \calM (\calK_\N)\mid f=0
 \text{ or } (f) \geq \calD\}, \;
r(\calD) =\dim_\bbC \calL (\calD),
\lb{a39}\\
\calL^1 (\calD) & =
 \{ \omega\in \calM^1 (\calK_\N)\mid \omega=0
 \text{ or } (\omega) \geq
\calD\},\; i(\calD) =\dim_\bbC \calL^1 (\calD),
\lb{a40}
\end{align}
($i(\calD)$ the index of speciality of $\calD$) one
infers that $\deg
(\calD)$, $r(\calD)$, and $i(\calD)$ only depend on
the divisor class
$[\calD]$ of $\calD$.  Moreover, we recall the
following fundamental
facts.

\begin{theorem} \lb{thm1}
Let $\calD \in \Div(\calK_\N)$,
$\omega \in \calM^1 (\calK_\N) \setminus \{0\}$. Then
\begin{equation}
 i(\calD) =r(\calD-(\omega)), \quad \N\in\bbN_0.
\lb{a41}
\end{equation}
The Riemann-Roch theorem reads
\begin{equation}
r(-\calD) =\deg (\calD) + i (\calD) -\N+1,
\quad \N\in\bbN_0.
\lb{a42}
\end{equation}
By Abel's theorem, $\calD\in \Div(\calK_\N)$,
$\N\in\bbN$ is principal
if and only if
\begin{equation}
\deg (\calD) =0 \text{ and } \ul \alpha_{Q_0} (\calD)
=\ul{0}.
\lb{a43}
\end{equation}
Finally, assume
$\N\in\bbN$. Then $\ul \alpha_{Q_0}
: \Div(\calK_\N) \to J(\calK_\N)$ is surjective
(Jacobi's inversion theorem).
\end{theorem}

\begin{lemma} \lb{la2}
Let $\calD_{\ul Q} \in \sigma^n \calK_\N$,
$\ul Q=(Q_1, \ldots, Q_\N)$.  Then
\begin{equation}
1 \leq i (\calD_{\ul Q} ) =s(\leq \N/2)
\lb{a46}
\end{equation}
if and only if there are $s$ pairs of the type
$(P, P^*)\in \{Q_1,
\ldots, Q_\N\}$ (this includes, of course, branch
points for which
$P=P^*$).
\end{lemma}

Denote by $\ul \Xi_{Q_0}=(\Xi_{Q_{0,1}}, \dots,
\Xi_{Q_{0,\N}})$ the vector of Riemann constants,
\begin{equation}
\Xi_{Q_{0,j}}=\frac12(1+\tau_{j,j})-
\sum_{\substack{\ell=1 \\ \ell\neq j}}^\N\int_{a_\ell}
\omega_\ell(P)\int_{Q_0}^P\omega_j,
\quad j=1,\dots,\N. \lb{aa55}
\end{equation}

\begin{theorem} \lb{taa17a}
Let $\ul Q =(Q_1,\dots,Q_\N)\in \sigma^\N \calK_\N$ and
assume $\calD_{\ul Q}$ to be nonspecial, that is,
$i(\calD_{\ul Q})=0$. Then
\begin{equation}
\theta(\ul {\Xi}_{Q_0} -\ul {A}_{Q_0}(P) + \alpha_{Q_0}
(\calD_{\ul Q}))=0 \text{ if and only if }
P\in\{Q_1,\dots,Q_\N\}. \lb{aa55a}
\end{equation}
\end{theorem}

\begin{remark} \lb{raa19}
While $\theta(\ul z)$ is well-defined (in fact, entire)
for $\ul z\in\bbC^\N$, it is not well-defined on
$J(\calK_\N)=\bbC^\N/L_\N$ because of  \eqref{aa51}.
Nevertheless, $\theta$ is a ``multiplicative
function'' on $J(\calK_\N)$ since the multipliers in
\eqref{aa51} cannot vanish.  In particular, if
$\ul z_1=\ul z_2\pmod{L_\N}$, then $\theta(\ul z_1)=0$
if and only
if  $\theta(\ul z_2)=0$.  Hence it is
meaningful to state that $\theta$ vanishes at points of
$J(\calK_\N)$. Since the Abel map
$\ul A_{Q_0}$ maps $\calK_\N$ into  $J(\calK_\N)$, the function
$\theta(\ul A_{Q_0}(P)-\ul{\xi})$ for
$\ul\xi\in\bbC^\N$, becomes a multiplicative function on
$\calK_\N$. Again it makes sense to say that
$\theta(\ul A_{Q_0}(\dott)-\ul{\xi})$ vanishes at points of
$\calK_\N$.
\end{remark}

Finally, we formulate the following auxiliary result
(cf., e.g., Lemma 3.4 in \cite{GR96}).

\begin{lemma} \lb{lem34}
Let $\psi(\dott,x),\; x\in\calU$, $\calU\subseteq\bbR$
open, be
meromorphic on ${\calK}_\N\setminus
\{\Pinf\}$ with an essential singularity
at $\Pinf$ such
that $\tilde \psi(\dott,x)$ defined by
\begin{equation}
\tilde \psi (\dott,x) =\psi (\dott,x)
\exp \bigg( -i(x-x_0) \int_{Q_0}^P
\Omega_{0}^{(2)}\bigg)
\lb{342a}
\end{equation}
is multi-valued meromorphic on ${\calK}_\N$ and its
divisor satisfies
\begin{equation}
(\tilde \psi (\dott,x))\geq -{\calD}_{\hat{\ul \mu}(x)}.
\lb{342b}
\end{equation}
Define a divisor ${\calD}_0 (x)$ by
\begin{equation}
(\tilde \psi (\dott,x))={\calD}_0 (x)
-{\calD}_{\hat{\ul \mu} (x)}.
\lb{342c}
\end{equation}
Then
\begin{equation}
{\calD}_0 (x) \in\sigma^\N {\calK}_\N, \; {\calD}_0 (x) \geq 0,
\; \deg ({\calD}_0 (x))
=\N.
\lb{342d}
\end{equation}
Moreover, if ${\calD}_0 (x)$ is nonspecial for all
$x\in\calU$, that is, if $i ({\calD}_0 (x) ) =0$,
then $\psi (\dott,x)$ is unique up to a constant
multiple (which may depend on $x$).
\end{lemma}

\section{Lagrange's interpolation formula and symmetric
functions}\lb{B}
\renewcommand{\theequation}{B.\arabic{equation}}
\renewcommand{\thetheorem}{B.\arabic{theorem}}
\setcounter{theorem}{0}
\setcounter{equation}{0}

In this appendix we quote a few useful facts from
\cite{EGHL97}, which facilitate the proofs of the
linearization properties of \eqref{3.23a} and \eqref{4.33c}.

Let $\N\in\bbN$ be fixed, introduce
\begin{subequations} \lb{ind}
\begin{align}
\mathcal S_k&=\{\ul\ell=(\ell_1,\dots,\ell_k)\in\bbN^k \mid
\ell_1<\cdots<\ell_k\leq \N \}, \quad k\le \N,\lb{inds} \\
\mathcal I^{(j)}_k&=\{\ul\ell=(\ell_1,\dots,
\ell_k)\in\mathcal S_k\mid
\ell_m\neq j\}, \quad k\le \N-1,  \lb{indt}
\end{align}
\end{subequations}
and define
\begin{subequations} \lb{funct}
\begin{align}
\Psi_{0}(\ul\mu)&=1, \quad \Psi_{k}(\ul\mu)=
(-1)^k\sum_{\ul\ell\in\mathcal
S_k}\mu_{\ell_1}\cdots\mu_{\ell_k},\quad k\le \N,
\lb{functpsi} \\
\Phi_0^{(j)}(\ul\mu)&=1, \quad
\Phi_k^{(j)}(\ul\mu)=(-1)^k\sum_{\ul\ell\in\mathcal
I^{(j)}_k}\mu_{\ell_1}\cdots\mu_{\ell_k},
\quad k\le \N-1, \lb{functphi} \\
\Phi_\N^{(j)}(\ul\mu)&=0, \no
\end{align}
\end{subequations}
where $\ul\mu=(\mu_{1},\dots,\mu_{\N})\in\bbC^\N$. We
recall our
notation
\begin{equation}
F_\N(z)=\prod_{j=1}^\N (z-\mu_j)
\end{equation}
and note
\begin{equation}
F_\N^\prime(\mu_j)=\prod_{\substack{\ell=1\\
\ell\neq j}}^\N (\mu_j-\mu_\ell)
\end{equation}
(the $(x,t_r)$-dependence of $\mu_j$ being
immaterial here and hence suppressed in the following).

\begin{theorem}[Lagrange's interpolation formula] \lb{theoremB1}
Assume that $\mu_1,\dots,\mu_\N$ are $\N$ distinct complex
numbers. Then
\begin{align}
&\sum_{j=1}^\N
\f{\mu_j^{m-1}}{F_\N^\prime(\mu_j)}\Phi_k^{(j)}(\ul\mu)
=\delta_{m,\N-k}-\Psi_{k+1}(\ul\mu)\delta_{m,\N+1}, \label{A1c} \\
& \hspace*{4cm} m=1,\dots, \N+1,\quad k=0,\dots, \N-1. \no
\end{align}
\end{theorem}
For a proof of Theorem \ref{theoremB1} see, for instance,
\cite{EGHL97}, Appendix A.

We will also need the following high-energy
expansion\footnote{Observe that the right-hand side
gives the homogeneous quantities $\hat f_j$ even with a
non-homogeneous left-hand side
$F_\N$.} (\cite{GD75}, \cite{GRT96})
\begin{equation}
\f{F_\N(z,x,t_r)}{y(P)}\underset{\zeta\to 0}{=}
\zeta\sum_{j=0}^\infty \hat f_j(x,t_r)\zeta^{2j},
\quad\zeta=\sigma /z^{1/2}, \quad \sigma =\pm 1 . \lb{high}
\end{equation}

Furthermore, we have\footnote{$n\mini m=\min\{n,m\}$.}
(cf.\ \cite{EGHL97},
Lemma 4.3)
\begin{equation}
\hat f_j=\sum_{k=0}^{j \mini \N} c_{j-k}(\ul E) \Psi_k(\ul\mu),
\lb{f-sym}
\end{equation}
where $\ul E =(E_0,\dots,E_{2\N})$
and\footnote{$(2n-1)!!=1\cdot3\cdots (2n-1)$, and
$(-1)!!=1$.}
\begin{align}
& c_0(\ul E)=1, \no \\
& c_k(\ul E)=\sum_{\substack{j_0,\dots,j_{2\N}=0\\
j_0+\cdots+j_{2\N}=k}}^{k}
\f{(2j_0-1)!!\cdots(2j_{2\N}-1)!!}
{2^k j_0!\cdots j_{2\N}!}E_0^{j_0}\cdots
E_{2\N}^{j_{2\N}}, \quad k\in\bbN. \lb{c}
\end{align}
In addition we recall\footnote{$n\maxi m=\max\{n,m\}$.} (see
\cite{EGHL97}, Lemma 4.4)
\begin{equation}
\hatt F_q(\mu_j)=\sum_{p=(q-\N)\maxi 0}^q c_p(\ul E)
\Phi_{q-p}^{(j)}(\ul\mu), \quad q=0,\dots,r-1. \lb{F-sym}
\end{equation}
We note that relations \eqref{f-sym} and \eqref{F-sym}
coincide with those
in the KdV case
since the recursion relations  are identical (but with the
KdV solution $V$
replaced by
$-(u_x^2+2 i u_{xx})/4$, see \eqref{2.13a}).
Observe the identity
\begin{equation}
\ti F_r(z)=z \ti F_{r-1}(z)+ \tilde f_r
\end{equation}
(using the same set of constants $c_{\ell}$,
$\ell=1,\dots,r-1$
in $\ti F_{r-1}$ and $\ti F_r$), which implies
\begin{equation}
\f{\ti F_r(\mu_j)}{\mu_j}=\ti F_{r-1}(\mu_j)
+\f{\tilde f_r}{\mu_j}=
\sum_{q=0}^{r-1} \tilde c_{r-1-q}
\sum_{p=(q-\N)\maxi 0}^q c_p(\ul E)
\Phi_{q-p}^{(j)}(\ul\mu)-\frac{\tilde
\alpha}{\alpha}\Phi_{\N-1}^{(j)}(\ul\mu), \lb{symm}
\end{equation}
using $\tilde f_r=\tilde \alpha e^{-iu}$ and the
trace relation
\eqref{4.33a}.

\section{Solvability of the recursion relation}\lb{C}
\renewcommand{\theequation}{C.\arabic{equation}}
\renewcommand{\thetheorem}{C.\arabic{theorem}}
\setcounter{theorem}{0}
\setcounter{equation}{0}

In this appendix we prove solvability
of the principal recursion relations in Section \ref{s2} and
show that the system of equations \eqref{2.23},
\eqref{2.24}, and
\eqref{2.26}--\eqref{2.28} is compatible.

We first prove that equations \eqref{2.23}, \eqref{2.26}, and
\eqref{2.28} are
consistent for all $j\in\bbN_0$, that is, without reference
to a fixed $\g\in\bbN_0$. To this end we define
$\{f_j\}_{j\in\bbN_0}$ by
\begin{align}
f^\pm_0 = 1,\quad f^\pm_{j,x} &= - \frac14 f^\pm_{j-1, xxx}
+ w_\pm
f^\pm_{j-1,x}
+ \frac12 w_{\pm,x} f^\pm_{j-1} \no \\
&=-\f14(\partial_x \mp iu_x)\partial_x(\partial_x\pm iu_x)
f^\pm_{j-1}, \quad
j\in\bbN,\lb{C1}
\end{align}
with integration constants $c^\pm_j\in\bbC$.  When choosing
the integration constants
equal, that is, $c^+_j=c^-_j = c_j$ for $j\in\bbN$, we
conclude that
\begin{equation}
f_j=f_j^+, \quad h_j=f_j^-, \quad j\in\bbN_0, \lb{C2}
\end{equation}
temporarily ignoring \eqref{2.24} and \eqref{2.27}. Observe
that $f^\pm_j$ are
defined by the same recursion used to construct
the quantity $f_j$ for the KdV hierarchy with the solution
$V$ replaced by
$w_\pm=-(u_x^2\pm 2i u_{xx})/4$.  It is  known that
$f^\pm_{j,x}$  equal a total derivative, and hence that
$f^\pm_{j}$ are differential
polynomials in $w_\pm$. Furthermore, let
\begin{equation}
g_{-1}^{\pm}=0, \quad g_j^\pm=\f{i}2(\pm f^\pm_{j,x}+
i u_x f^\pm_{j})=\pm
\f{i}2 (\partial_x\pm
iu_x) f^\pm_{j},
\quad  j\in\bbN_0.
\lb{C3}
\end{equation}
Combining \eqref{C1} and \eqref{C3} yields
\begin{equation}
f^\pm_{j,x}=\pm\f{i}2(\partial_x \mp iu_x) g^\pm_{j-1,x},
\quad j\in\bbN_0.
\lb{C4}
\end{equation}
Write
\begin{equation}
f^\pm_{j}=\pm \f{i}2 g^\pm_{j-1,x}+\rho_j^\pm+c^\pm_j, \lb{C4aa}
\end{equation}
where $\rho_j^\pm$ is a differential polynomial in $w_\pm$ with
\begin{equation}
\rho^\pm_{j,x}=f^\pm_{j,x}\mp \f{i}2 g^\pm_{j-1,xx}=
\f12 u_x g^\pm_{j-1,x}, \lb{C4ab}
\end{equation}
using \eqref{C4}.
\begin{lemma}\lb{lemmaC1}
Let $f^\pm_{j}$ and $g_j^\pm$ be defined by \eqref{C1}
and \eqref{C3},
respectively.  Assume that
\begin{equation}
c^+_j=c^-_j = c_j, \quad j\in\bbN. \lb{C1a}
\end{equation}
Then
\begin{equation}
g_j^+=g_j^- = g_j, \quad j\in\bbN_0, \lb{C4a}
\end{equation}
where $g_j$ is defined as in \eqref{2.23} or \eqref{2.26}.
Moreover,
\begin{equation}
g_{j,x}=i(f_{j+1}^- -f_{j+1}^+), \quad j\in\bbN_0. \lb{C5}
\end{equation}
\end{lemma}
\begin{proof}
Equation \eqref{C4a} is certainly
true for $j=0$. Proceeding inductively upon $j$ we assume
that $g_{j-1}^+=g_{j-1}^- = g_{j-1}$ holds. Then
\begin{equation}
f^\pm_{j,x} =\pm\f{i}2(\partial_x \mp iu_x) g_{j-1,x}. \lb{C6}
\end{equation}
Using this, \eqref{C4}, \eqref{C4aa}, and \eqref{C4ab} we
conclude that
$\rho_j^+=\rho_j^-=\rho_j$ and hence
\begin{equation}
f^\pm_{j} =\pm\f{i}2 g_{j-1,xx}+\rho_j+c^\pm_j, \lb{C7}
\end{equation}
with $\rho_{j,x}=u_x g_{j-1,x}/2$. Consequently,
\begin{align}
g_j^+-g_j^-&=\f{i}2 (\partial_x+ iu_x) f^+_{j}+
\f{i}2 (\partial_x- iu_x)
f^-_{j} \no \\
&=\f{i}2\big((\partial_x+ iu_x)(\f{i}2 g_{j-1,x}
+\rho_j+c^+_j)
+ (\partial_x - iu_x)(-\f{i}2 g_{j-1,x}+
\rho_j+c^-_j)\big) \no \\
&=\f12 u_x(c^+_j-c^-_j)-\f{i}2 u_x g_{j-1,x}+i\rho_{j,x}
=0 \lb{C8}
\end{align}
using \eqref{C1a}. This proves \eqref{C4a}.

By \eqref{C3}--\eqref{C4a} one infers
\begin{align}
0&=g^+_{j+1} - g^-_{j+1} =\frac{i}{2}(f^+_{j+1,x}+f^-_{j+1,x})
-\frac{1}{2}u_x(f^+_{j+1}-f^-_{j+1}) \no \\
&=\frac{i}{2}\big(\frac{i}{2}(\partial_x-iu_x)g_{j,x}
-\frac{i}{2}(\partial_x+iu_x)g_{j,x}\big)
-\frac{1}{2}u_x(f^+_{j+1}-f^-_{j+1}) \no \\
&=\frac{i}{2}u_x(g_{j,x}-i(f^-_{j+1}-f^+_{j+1})) \lb{C9}
\end{align}
and hence \eqref{C5} as long as $u_x \not\equiv 0$. The latter
case can only occur for $\N=0$ but then \eqref{C5} can
easily be verified directly.
\end{proof}
At this point we proved that equations \eqref{2.23},
\eqref{2.26} and
\eqref{2.28} are
compatible.  To satisfy the additional equations
\eqref{2.24}, \eqref{2.27} and hence
\eqref{2.30}, we proceed as follows. Fix $\g\in\bbN$ and  let
$\alpha, \beta\in\bbC$. In accordance with the notation
used in this
appendix we
write $\alpha=\alpha^+$ and $\beta=\alpha^-$. We define the
$\g$th stationary sGmKdV equation by the relation
\eqref{2.30}, \eqref{2.26a}, that is,
\begin{equation}
g_{\g-1,x}=i(\alpha^- e^{iu}-\alpha^+ e^{-iu}),
\quad \N\in\bbN_0, \lb{C10}
\end{equation}
subject to the constraint
\begin{equation}
f_\N^+(x)f_\N^-(x)=\alpha^+ \alpha^- =P_{2\N}(0). \lb{C10a}
\end{equation}
\begin{lemma} \lb{lemmaC2}
Let $f_j^{\pm}$ and $g_j$ be defined according to \eqref{C1},
\eqref{C1a}, \eqref{C3}, \eqref{C4a} and assume \eqref{C10} and
\eqref{C10a}. Then
\begin{equation}
f^\pm_{\g}= \alpha^\pm e^{\mp iu}, \quad \N\in\bbN. \lb{C11}
\end{equation}
\end{lemma}
\begin{proof}
Using \eqref{C6} for $j=\g$ one finds
\begin{align}
f^\pm_{\g,x} &=\pm\f{i}2(\partial_x \mp iu_x) g_{\g-1,x}
=\pm\f{i}2 g_{\g-1,xx}+ \f12 u_x g_{\g-1,x} \no \\
&=\mp \f{1}2 \partial_x(\alpha^- e^{iu}-\alpha^+
e^{-iu}) + \f12 u_x
g_{\g-1,x} \no \\
&=\mp \f{i}2 (\alpha^- e^{iu}+\alpha^+ e^{-iu})u_x
 + \f12u_x g_{\g-1,x} \no \\
&=\mp \f{i}2 (\alpha^- e^{iu}-\alpha^+
e^{-iu})u_x \mp i \alpha^\pm e^{\mp
iu} u_x
 + \f12 u_x g_{\g-1,x} \no \\
&=-\f12 u_x g_{j-1,x}\mp i \alpha^\pm e^{\mp iu} u_x
                + \f12 u_x g_{\g-1,x}
=\mp i \alpha^\pm e^{\mp iu} u_x \no \\
&=(\alpha^{\pm} e^{\mp iu})_x, \lb{C12}
\end{align}
using relation \eqref{C10} twice. Thus
\begin{equation}
f_\N^{\pm}(x)=\alpha^{\pm} e^{\mp iu(x)} + \gamma^{\pm} \lb{C13}
\end{equation}
for some constants $\gamma^{\pm}\in\bbC$. By \eqref{C10a},
\begin{equation}
P_{2\N}(0)=f_\N^+(x)f_\N^-(x)=\alpha^+\alpha^-
=\alpha^+\alpha^-+\gamma^+\gamma^-
+\alpha^+\gamma^-e^{-iu(x)} + \alpha^-\gamma^+e^{iu(x)} \lb{C14}
\end{equation}
then yields $\gamma^+=\gamma^-=0$ and hence \eqref{C11}.
\end{proof}

The case $\N=0$ yields the trivial situation $u_x \equiv 0$,
$f^+_0f^-_0=P_0(0)$, $f^+_0=f^-_0$ by \eqref{2.4},
\eqref{2.6}, \eqref{2.28k}, and \eqref{2.30}.

{\bf Acknowledgments.} H.H. is indebted to the Department of
Mathematics at the University of
Missouri, Columbia for the great hospitality
extended to him during his sabbatical 1996--97 when this
work was done.
Financial support by the Norwegian Research Council is
gratefully acknowledged.


\end{document}